\documentclass[journal,twoside,web]{IEEEtran}
\usepackage{cite}
\usepackage{amsmath,amssymb,amsfonts}
\usepackage{graphicx}
\usepackage{textcomp}
\usepackage{graphicx}          
\usepackage{bm}
\usepackage{amsmath}
\usepackage{graphicx}
\usepackage{amssymb}
\usepackage{multirow}
\usepackage{array}
\usepackage{color}
\usepackage{epsfig}
\usepackage{graphics} 
\usepackage{algorithm}
\usepackage{algorithmicx}
\usepackage{epsfig} 
\usepackage[noend]{algpseudocode}

\allowdisplaybreaks

\newcommand{\lfteqn}{\begin{eqnarray} \begin{array}{lllllll}}
		\newcommand{\ndeqn}{\end{array} \nonumber \end{eqnarray}}
\newcommand{\Lfteqn}{\begin{eqnarray} \begin{array}{lllllll}}
		\newcommand{\Ndeqn}{\end{array}  \end{eqnarray}}

\newtheorem{theorem}{Theorem}

\newtheorem{definition}{Definition}
\newtheorem{example}{Example}
\newtheorem{proposition}{Proposition}
\newtheorem{remark}{Remark}

\newcommand{\cS}{\mathcal{S}}

\newcommand{\bed}{\begin{displaymath}}
	\newcommand{\eed}{\end{displaymath}}
\newcommand{\bea}{\bed\begin{array}{rl}}
	\newcommand{\eea}{\end{array}\eed}
\newcommand{\beq}[1]{\begin{equation} \label{#1}}
	\newcommand{\eeq}{\end{equation}}
\newcommand{\barray}{\begin{array}{ll}}
	\newcommand{\earray}{\end{array}}

\makeatletter
\newenvironment{breakablealgorithm}
{
	\begin{center}
		\refstepcounter{algorithm}
		\hrule height.8pt depth0pt \kern2pt
		\renewcommand{\caption}[2][\relax]{
			{\raggedright\textbf{\ALG@name~\thealgorithm} ##2\par}%
			\ifx\relax##1\relax 
			\addcontentsline{loa}{algorithm}{\protect\numberline{\thealgorithm}##2}%
			\else 
			\addcontentsline{loa}{algorithm}{\protect\numberline{\thealgorithm}##1}%
			\fi
			\kern2pt\hrule\kern2pt
		}
	}{
		\kern2pt\hrule\relax
	\end{center}
}
\makeatother

\begin{document}

\title{Modeling and Control of Discrete Event Systems under
	Joint Sensor-Actuator Cyber Attacks}
\author{Shengbao Zheng, Shaolong Shu,\IEEEmembership{Senior Member, IEEE} and Feng Lin, \IEEEmembership{Fellow, IEEE}
	\thanks{The authors of this paper are supported by the National Natural Science Foundation of China under Grants 61773287, 62073242, and by the Shanghai Municipal Science and Technology Major Project (2021SHZDZX0100) and the Fundamental Research Funds for the Central Universities. }
	\thanks{S. Zheng and S. Shu are with the School of Electronics and Information Engineering, Tongji University, Shanghai 200072, China (e-mail:,
		1910640@tongji.edu.cn; shushaolong@tongji.edu.cn).}
	\thanks{F. Lin is with the School of Electronics and Information Engineering,
		Tongji University, Shanghai 200072, China, and also with the Department of Electrical and Computer Engineering, Wayne State University,
		Detroit, MI 48202 USA (e-mail:,flin@wayne.edu).}}

\maketitle

\begin{abstract}
	In this paper, we investigate joint sensor-actuator cyber attacks
	in discrete event systems. We assume that attackers can attack
	some sensors and actuators at the same time by altering
	observations and control commands. Because of the nondeterminism
	in observation and control caused by cyber attacks, the behavior
	of the supervised system becomes nondeterministic and may deviate
	from the safety specification. We define the upper-bound on all possible languages that can be generated by the supervised system to investigate the safety supervisory control problem under cyber attacks. After introducing CA-controllability and CA-observability, 
	we prove that the supervisory control problem under cyber attacks is solvable if and only if the given specification language is CA-controllable and CA-observable. Furthermore, we obtain methods to calculate the state estimates under sensor attacks and to synthesize a state-estimate-based supervisor to achieve a given safety
	specification under cyber attacks. We further show that of all the solutions, the proposed state-estimate-based supervisor is maximally-permissive.
	
\end{abstract}


\section{Introduction}

In the past ten years and more, cyber-physical systems have become a hot research topic \cite{Baheti,7815549}. For a typical cyber-physical system, the information is transmitted via wired/wireless networks and such networked systems are vulnerable to 
cyber attacks \cite{Cardenas,lin2014control}. Indeed, several attacks on networked systems have been reported in the literature \cite{Dibaji,6545301,6096958,5530638}; the following examples are given in \cite{Dibaji,Stuxnet,RQ170,2013Cyber,Maroochy}: (1) Stuxnet attack on an Iranian uranium enrichment plant in 2011; (2) RQ-170 attack in 2011, where US operators lost control of an RQ-170 unmanned aerial vehicle which subsequently landed in Iran; 
(3) a series of attacks on Ukrainian power distribution networks that caused outages as well as lasting damage in 2015; (4) the Maroochy water services attack in Queensland, Australia by a disgruntled employee in 2000; and (5) a Jeep hijacked remotely by attackers while driving at 70 mph on a highway in St. Louis, USA.

Cyber attacks may change the readings of sensors and hence change
the observations of controllers which will cause the controllers
to make incorrect decisions. They may also alter the actions of
actuators and hence change the behavior of the system.
How to protect the feedback control systems from cyber attacks in
critical infrastructures has become an increasingly important
problem \cite{Dibaji,Jin,Antsaklis,rashidinejad2019supervisory}.

At the supervisory layer, a cyber-physical system can be abstracted as a discrete event system and the control objective is often to ensure that the supervised system is safe \cite{cassandras2009introduction, ramadge1987supervisory, lin1988observability}. Within the discrete event system framework, cyber attacks are investigated extensively. 

An early work is done in \cite{Thorsley2006} which shows that cyber attacks can damage the security of the supervised system and proposes a method to estimate how much damage that cyber attacks might cause. In \cite{Carvalho2016, Carvalho2018}, four types of cyber attacks are discussed: event insertion attacks, event
deletion attacks, event enablement attacks  and event disablement
attacks. In these works, a module is embedded in the supervisor to detect cyber attacks. Once an attack is detected, the supervisor will disable all controllable events. 

If cyber attacks cannot be detected in time, the above methods
cannot ensure the safety/security of the supervised systems. Hence
more work has been done to investigate  deception attacks \cite{zhang2021joint,romulr2017,romulr2019s,Meira-Goesp,Meira-Goespj,Sahar}. Some work considers how to synthesize deception attack 
strategies. In \cite{romulr2017}, the authors show how to synthesize a deterministic attack strategy. \cite{romulr2019s} proposes a method to synthesize a non-deterministic attack strategy. 
In \cite{Sahar}, an equivalence
relation is proposed to reduce the complexity of synthesizing a
deception attack strategy. On the other hand, some work considers
how to design a robust supervisor against deception attacks. In
\cite{Surong}, the authors propose a supervisor against bounded
sensor attacks. In \cite{wakaki},  the supervisory control
problem under deception attacks is solved and the necessary
and sufficient conditions are derived. 
In \cite{romulotowards}, a robust
supervisor is synthesized to mitigate the attacks on the sensor
reading based on the solution of a partially observed supervisory
control problem with arbitrary control patterns. Using game
theory, \cite{romulr2021} proposes a bipartite-graph-based method
to solve the supervisory control problem for partially-observed
discrete event systems under cyber attacks. All these works focus
on sensor attacks.

For actuator attacks, \cite{normal} proposes a method to
synthesize a deception attack strategy through subset
construction. In \cite{linforfree}, the synthesis problem of
deception attack strategies is converted into a traditional
supervisor synthesis problem. Based on the attack model in
\cite{normal}, an obfuscated supervisor is proposed in
\cite{zhuyu} to prevent the supervised system from entering unsafe
states even though actuator attacks may occur using a technique
of solving the boolean satisfiability problem.

In practice, sensors and actuators are usually distributed in the same network in a cyber-physical system. The attackers can invade the sensors to change their readings and invade the actuators to alter their actions at the same time. In this sense, we need to consider the impacts of joint sensor and actuator attacks on the supervised systems 
\cite{lima2017security,Lima2018,Lima2019,lima2021security,lin2019towards,wang2019attack,lin2020synthesis,lin2021synthesis}. \cite{lima2017security,Lima2018,Lima2019,lima2021security} call the joint sensor and actuator attacks as man-in-the-middle attacks. They assume sensor attacks can delete the current observed event or replace it with another observable event and the actuator attacks can enable a disabled event or disable an enabled event in the attackable event set. 
The control is to ensure that the system runs normally before attacks are detected and not enters unsafe states after attacks are detected. A security supervisor is synthesized to control the given system together with the normal supervisor. The security supervisor begins to work when attacks are detected. NA-security is proposed to ensure this control problem has solutions. \cite{lin2019towards} uses the similar models for sensor attacks and actuator attacks. The control object is to ensure the dynamics of the closed-loop control system is in a given range and can be solved with the range supervisory control framework. In \cite{wang2019attack}, sensor attacks are nondeterministic and the attacker can randomly choose one observation/control from a set to replace the current observation/control. In this case, the observation and control for an occurred string becomes nondeterministic and hence the dynamics of the closed-loop control system becomes nondeterministic. A supervisor is synthesized which ensures the closed-loop control system is deterministic (a string can either occur in all possible attacks or never occur in any possible attack), which may be hard to achieve. In \cite{lin2020synthesis,lin2021synthesis}, the authors investigate how to find a powerful joint sensor and actuator attack policy, not how to synthesize a supervisor to tame attacks. 

As shown in the above papers, the conventional supervisory control theory cannot handle cyber attacks in the sense that the supervisor designed cannot ensure safety of the supervised system. In order to design a robust supervisor that ensures safety of the supervised system under a large class of joint sensor-actuator cyber attacks, we develop a new supervisory control theory. The theory uses a new cyber attack model that is more general than the existing models in the sense that the existing models are special cases of the general model. More specifically, we assume that some observable events are vulnerable to cyber attacks.  
When a vulnerable observable event occurred, we assume its observation may be
changed to a string in a language described by an automaton. It means the attacker can randomly choose one string in the language to replace the occurred vulnerable observable event. 
We further assume some controllable events are vulnerable to cyber
attacks. For a vulnerable controllable event, no matter what is the control command issued by the supervisor, the attackers can disable or enable it. Because of the nondeterminism in observation and
control caused by cyber attacks, the behavior of the supervised
systems becomes nondeterministic. Inspired by previous results on networked control of discrete event systems \cite{lin2014control, ShuSupervisor, ShuDeterministic, ShuPredictive}, we define an upper-bound on the closed-loop languages. We then use the
upper-bound language to investigate the safety supervisory control
problem under cyber attacks. In order to obtain necessary and
sufficient conditions for the existence of solutions, we introduce two new concepts, CA-controllability and CA-observability. We prove that there exists a supervisor under which the upper-bound language of the supervised system under cyber attacks is equal to the specification language if and only if the specification language is CA-controllable and CA-observable. We further synthesize a state-estimate-based supervisor by developing a systematic method to calculate state estimates for observations subject to cyber attacks. The proposed state-estimate-based supervisor is optimal in the sense that it is maximally-permissive among all the valid supervisors (solutions to the safety supervisory control problem under cyber attacks).

The main differences between this paper and papers mentioned above are as follows. 
(1) A more general model for sensor attacks is proposed in this paper. The model is more general in the sense that other models such as models for deletions, insertions, and replacements are special cases of the general model. 
(2) In order to deal with nondeterminism in a supervised system caused by nondeterministic attacks, an upper-bound language of the supervised system under cyber attacks is defined. The control object is to ensure the safety of the supervised system by requiring that the upper-bound language is equal to or contained in a given specification language $K$.
(3) A new supervisory control theory is developed and necessary and sufficient conditions are derived for the existence of a supervisor under cyber attacks whose upper-bound language is equal to $K$. The conditions are given in terms of two new concepts of CA-controllability and CA-observability.
(4) Using the general model and the new theory, new methods are proposed to calculate state estimates and to design a robust supervisor that works for all cyber attacks allowed by the model. 

The paper is organized as follows. Section \ref{s2} introduces
discrete event systems and cyber attacks. Section \ref{s3}
formally states the safety supervisory control problem under cyber
attacks. Section \ref{s4} proposes a method to calculate state
estimates from observations under cyber attacks and a
state-estimate-based supervisor is then proposed. Section \ref{s5}
solves the supervisory control problem under cyber attacks.
Section \ref{s6} finds the maximally-permissive supervisor for the
supervisory control problem under cyber attacks. In Section
\ref{s7}, we extend the results to a more general case where
sensor attack strategies are observation-based. Finally, we
conclude the paper in Section \ref{s8}. Preliminary version of some results
on the supervisory control problem under cyber attacks in this paper is presented in \cite{zheng2021modeling}. Compared with \cite{zheng2021modeling}, this paper contains proofs omitted in \cite{zheng2021modeling} as well as
detailed explanations and examples. We further
investigate the maximally-permissive supervisor and observation-based sensor attacks which are not considered in \cite{zheng2021modeling}.

\section{Discrete event systems under cyber attacks}\label{s2}

A discrete event system is modeled as a deterministic automaton
\cite{cassandras2009introduction}
$$
G=( Q, \Sigma , \delta , q_0 ),
$$
where $Q$ is the set of states; $\Sigma$ is the set of events;
$\delta : Q \times \Sigma \rightarrow Q$ is the (partial)
transition function; and $q_0$ is the initial state. The set of
all possible transitions is also denoted by $\delta$, that is,
$\delta = \{ (q, \sigma, q'): \delta (q, \sigma) =q' \}$.

We use $\Sigma ^*$ to denote the set of all strings over $\Sigma$.
The transition function $\delta$ can be extended to strings, that
is, $\delta : Q \times \Sigma^* \rightarrow Q$, in the usual way.
We use $\delta (q,s)!$ to denote that $\delta (q,s)$ is defined.
The language {\em generated} by $G$ is the set of all strings
defined in $G$ from the initial state $q_0$ as
$$
L(G) = \{ s \in \Sigma ^*: \delta (q_0,s)! \}.
$$

In general, a language $K \subseteq \Sigma ^*$ is a set of
strings.  For a string $s \in \Sigma ^*$, the set of all prefixes of $s$ is denoted as 
	$Pr(s)=\{s' \in \Sigma^*: \exists t \in \Sigma^*, s't=s\}$. The (prefix) closure of $K$, denoted as
	$\overline{K}$, is the set of prefixes of strings in $K$ as
	$$
	\overline{K}=\{s':(\exists s\in K)s' \in Pr(s)\}
	$$
	 A language is (prefix)
closed if it equals its prefix closure. By the definition, $L(G)$
is closed. For a string $s \in \Sigma ^*$, we use $|s|$ to denote
its length. For a set $x \subseteq Q$, we use $|x|$ to denote its
cardinality (the number of its elements).

We use a controller, called supervisor, to control the plant so
that some control objective is achieved. The supervisor can
control some events and observe some other events. The set of
events that can be controlled/disabled, called controllable
events, is denoted by $\Sigma _c$ ($\subseteq \Sigma$). $\Sigma
_{uc} = \Sigma - \Sigma _c$ is the set of uncontrollable events.
The set of events that can be observed, called observable events,
is denoted by $\Sigma _o$ ($\subseteq \Sigma$). $\Sigma _{uo} =
\Sigma - \Sigma _o$ is the set of unobservable events. The set of
observable transitions is denoted by $\delta_o = \{ (q, \sigma,
q')\in \delta : \sigma \in \Sigma _o\}$; and the set of
unobservable transitions is denoted by $\delta_{uo} = \{ (q,
\sigma, q') \in \delta : \sigma \in \Sigma _{uo}\}$.

For a given string, its observation is described by the natural
mapping $P:\Sigma^* \rightarrow \Sigma^*_o$ which is defined as
\begin{equation*}
	\begin{aligned}
		& P(\varepsilon)= \varepsilon\\
		& P(\sigma) =\begin{cases}
			\sigma &\text{if $\sigma \in \Sigma_o$}\\
			\varepsilon &\text{if $\sigma\in \Sigma - \Sigma_o$}\\
		\end{cases}\\
		& P(s\sigma)=P(s)P(\sigma) ,  s\in \Sigma^*, \sigma \in \Sigma,
	\end{aligned}
\end{equation*}
where $\varepsilon$ is the empty string.

As in \cite{Carvalho2018}, cyber attackers can change the readings of some related sensors. Hence, some observable events can be attacked. Denote the set of observable events and transitions that may be attacked by $\Sigma _o^a \subseteq \Sigma _o$ and $\delta ^a = \{ (q, \sigma, q') \in \delta : \sigma \in \Sigma _o^a \}$, respectively.

For a given transition $tr=(q, \sigma, q')\in \delta ^a$, we
assume that an attacker can change the observation from  event $\sigma$ to any
string in a language $A_{tr} \subseteq \Sigma_o ^*$. $A_{tr}$ is determined based the knowledge of the attacker and the corresponding sensor. We denote the
set of all such languages as $\mathbb{A}$
$$
\mathbb{A}=\{A_{tr}:tr=(q, \sigma, q')\in \delta ^a\}
$$
Sensor attack strategies are then described by a mapping from the
set of transitions to be attacked to the set of languages
$\mathbb{A}$
$$
\pi : \delta ^a\rightarrow \mathbb{A}
$$
where $\pi(tr)=A_{tr}$. Note that $\pi$ is transition-based. Note
also that this general definition allows for nondeterministic
attacks and includes the following special cases. (1) No attack:
if $\sigma \in A_{tr}$ and $\sigma$ is altered to $\sigma$ (no
change), then there is no attack. (2) Deletion: if the empty
string $\varepsilon \in A_{tr}$ and $\sigma$ is altered to
$\varepsilon$, then $\sigma$ is deleted. (3) Replacement: if
$\alpha \in A_{tr}$ and $\sigma$ is altered to $\alpha$, then
$\sigma$ is replaced by $\alpha$. (4) Insertion: if $\sigma \alpha
\in A_{tr}$ and $\sigma$ is altered to $\sigma \alpha$, then
$\alpha$ is inserted.


If a string $s = \sigma _1 \sigma _2 ..., \sigma _{|s|} \in L(G)$
occurs in $G$, the set of possible strings after cyber attacks in observation channel, denoted
by $\Theta ^\pi(s)$, is obtained as follows. Denote $q_k =
\delta(q_0,\sigma _1\cdots\sigma _{k}), k= 1, 2, ..., |s|$, then
$$
\Theta ^\pi(s) = L_1 L_2 ... L_{|s|},
$$
where
\begin{equation} \label{Equation1}
	\begin{split}
		L_k = \left\{ \begin{array}{ll}
			\{ \sigma _k \} & \mbox{if } (q_{k-1}, \sigma _k, q_k) \not\in \delta ^a \\
			A_{(q_{k-1}, \sigma _k, q_k)} & \mbox{if } (q_{k-1}, \sigma _k,
			q_k) \in \delta ^a \end{array} \right .
	\end{split}
\end{equation}
Note that $\Theta ^\pi(s)$ contains more than one string. Hence,
$\Theta ^\pi$ is a mapping from $L(G)$ to $2^{\Sigma ^*}$:
$$
\Theta ^\pi: L(G) \rightarrow 2^{\Sigma ^*}.
$$

The observation under both partial observation and cyber attacks in observation channel is then given by
$$
\Phi ^\pi = P \circ \Theta ^\pi,
$$
where $\circ$ denotes composition (of functions). In other words, for $s \in L(G)$, $\Phi ^\pi (s) = P ( \Theta ^\pi (s))$.
Hence, $\Phi ^\pi$ is a mapping from $L(G)$ to $2^{\Sigma _o^*}$:
$$
\Phi ^\pi: L(G) \rightarrow 2^{\Sigma _o ^*}.
$$

We extend $P$, $\Theta ^\pi$, and $\Phi ^\pi$ from strings $s$ to
languages $L$ in the usual way as
\begin{equation} \label{Equation5}
	\begin{split}
		& P(L) = \{ t \in \Sigma^*_o: (\exists s \in L) t = P(s) \} \\
		& \Theta ^\pi(L) = \{ t \in \Sigma^*: (\exists s \in L) t  \in
		\Theta ^\pi(s) \} \\
		& \Phi ^\pi(L) = \{ t \in \Sigma^*_o: (\exists s \in L) t  \in \Phi
		^\pi(s) \}
	\end{split}
\end{equation}

After the occurrence of $s\in L(G)$, a supervisor $\mathcal{S}$
observes one of the string $t \in \Phi ^\pi (s)$. Based on $t$,
$\mathcal{S}$ enables a set of events, denoted by
$\mathcal{S}(t)$. Hence, $\cS$ is a mapping
$$
\mathcal{S} :\Phi ^\pi ( L(G)) \rightarrow \Gamma ,
$$
where $\Gamma = 2^{\Sigma}$ is the set of all possible controls.
Note that $\Sigma_{uc}\subseteq \mathcal{S}(t)$ because
uncontrollable events are always permitted to occur.

We assume that the disablements/enablements of some controllable
events can be altered by attackers in the control channel in the
sense that an attacker can enable an event that is disabled by the
supervisor or disable an event that is enabled by the supervisor.
Denote the set of controllable events that may be attacked by
$\Sigma _c^a \subseteq \Sigma _c$. Note that uncontrollable events
are always permitted to occur and the attackers cannot disable
them.

Under cyber attacks in the control channel, for a given control
$\gamma \in \Gamma$, some events in $\Sigma _c^a$ can be added to it or removed from it by attackers. Hence the possible controls are
\begin{equation} \label{Equation11}
	\begin{split}
		\Delta(\gamma) = \{ \gamma_a \in 2^{\Sigma} :  (\exists \gamma ', \gamma ''
		\subseteq  \Sigma _c^a ) \gamma_a = (\gamma - \gamma ')\cup  \gamma ''\} .
	\end{split}
\end{equation}

When the supervisor issues a control command $\mathcal{S}(t)$
after observing $t \in \Phi ^\pi ( L(G))$, it may be altered under
cyber attacks. We use $\cS^a (t)$ to denote the set of all
possible control commands that may be received by the plant under
cyber attacks, that is,
\begin{equation} \label{Equation12}
		\cS^a(t)=\Delta(\cS(t)).
\end{equation}

Let us use an example to illustrate how cyber attacks change the
observations and controls.
\begin{example}\label{G} \rm
	We consider a discrete event system $G$ shown in Fig.
	\ref{attacks}. We assume that all the events are controllable and
	observable. That is, $\Sigma_c=\Sigma_o=\Sigma$. Observations of
	events $\lambda$ and $\mu$ can be changed by attackers, that is, 
	$\Sigma^a_o=\{\lambda, \mu\}$. Attackers can also enable/disable
	the occurrence of events $\alpha$ and $\beta$, that is, 
	$\Sigma^a_c=\{\alpha, \beta\}$. The observations of transition
	$tr_1=(2,\lambda, 3)$ under cyber attacks are described by
	$A_{tr_1}=\{\varepsilon,\lambda,  \lambda\mu\}$. The observations
	of $tr_2=(3, \mu, 1)$ under attacks are described by $A_{tr_2}=\{\mu,
	\beta\}$.
	
	\begin{figure}[htb]
		\centering
		\includegraphics[scale=1]{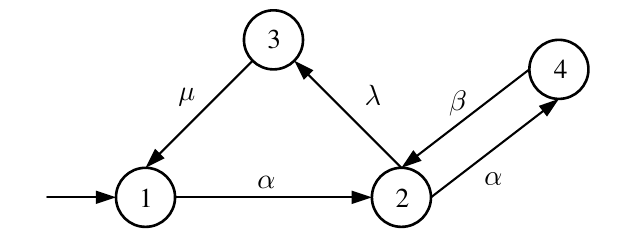}
		\caption{A discrete event system $G$.}
		\label{attacks}
	\end{figure}
	
	Let us consider the possible observations for string
	$s=\alpha\lambda$. By the definition, we have
	$$
	\Theta^\pi(s)=\{\alpha\} A_{tr_1}=\{\alpha, \alpha\lambda , \alpha\lambda\mu\}.
	$$
	We then have $\Phi^\pi(s)=P(\Theta^\pi(s))=\Theta^\pi(s)$  because all the
	events are observable.
	
	Let us consider another string $s'=\alpha\lambda\mu$. Its
	observations are
	\begin{equation*}
		\begin{aligned}
			\Phi^\pi(s')&=\Theta^\pi(\alpha\lambda\mu)=\{\alpha\}A_{tr_1}A_{tr_2}\\&=\{\alpha\mu, \alpha\beta, \alpha\lambda\mu , \alpha\lambda\beta, \alpha\lambda\mu\mu, \alpha\lambda\mu\beta\}
		\end{aligned}
	\end{equation*}
	
	If $\alpha\lambda\mu$ is observed, the supervisor issues a control $\cS(\alpha\lambda\mu)=\{\alpha, \lambda, \mu\}$. However, the actual control
	command received by $G$ is one of the following
	$$
	\Delta(\cS(\alpha\lambda\mu))=\{\{\alpha, \lambda, \mu\},
	\{\alpha, \beta, \lambda, \mu\}, \{\lambda, \mu\}\}.
	$$
Note that $\alpha$ is enabled by the supervisor without attack. However, if the control after attack is $\{\lambda, \mu\}$, then $\alpha$ is
disabled.\label{attackmodel}
	
\end{example}

\section{Problem statement}\label{s3}
\begin{figure}[htb]
	\centering
	\includegraphics[scale=1]{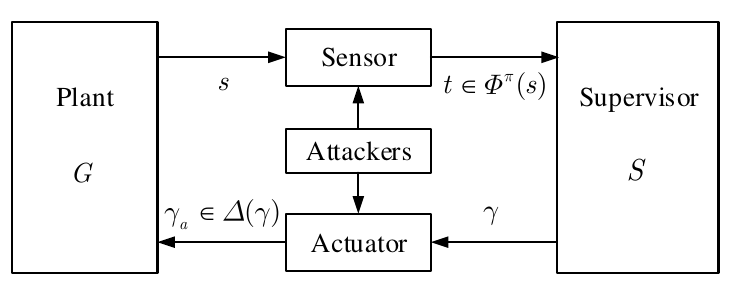}
	\caption{The closed-loop control framework under cyber attacks.}
	\label{closed-loop}
\end{figure}

As discussed in the previous section, when we use a supervisor to
control a discrete event system, both the observation and the
control command may be altered due to cyber attacks as shown in
Fig. \ref{closed-loop}.

Specifically, cyber attacks alter the observation of string $s$
from $P(s)$ to one in $\Phi ^\pi (s)$. They also alter the control
command from $\mathcal{S}(t)(=\gamma)$ to one in $\cS^a(t)=\Delta(\cS(t))$
for a given observation $t$.

We denote the supervised system under attacks as $\mathcal{S}
^a/G$. Because of the attacks, the language generated by the
supervised system, denoted by $L(\mathcal{S} ^a/G)$, is nondeterministic, and
hence may not be unique. We consider the upper
bound of all languages that can be generated by the supervised
system $\mathcal{S} ^a/G$. We call the upper bound {\em large language}, which is
given below.

\begin{definition} \label{Definition1} \rm
	The large language generated by the supervised system 		$L(\mathcal{S} ^a/G)$, denoted by $L_a (\mathcal{S} ^a/G)$, is 		defined recursively as follows.
	
	\begin{enumerate}
		\item
		The empty string belongs to $L_a (\cS ^a/G)$. That is,
		$$
		\varepsilon \in L_a (\cS ^a/G).
		$$
		\item
		If $s$ belongs to $L_a (\cS ^a/G)$, then for any $\sigma \in
		\Sigma$, $s \sigma$ belongs to $L_a (\cS ^a/G)$ if and only if $s
		\sigma$ is allowed in $L(G)$ and $\sigma$ is uncontrollable or
		enabled by $\cS ^a$ in {\em some} situations. That is, for any
		$ s \in L_a (\cS ^a/G)$ and $\sigma \in \Sigma$,
		\begin{equation} \label{Equation9}
			\begin{split}
				&  s \sigma\in L_a (\cS ^a/G) \\
				\Leftrightarrow & s \sigma \in L(G) \wedge (\sigma \in \Sigma
				_{uc} \\
				& \vee (\exists t \in \Phi ^\pi (s))(\exists \gamma_a \in \cS ^a(t))
				\sigma \in \gamma_a).
			\end{split}
		\end{equation}
		
	\end{enumerate}
	
\end{definition}

\begin{remark}
	Note that $L_a(\mathcal{S}^a/G)$ may include some strings which can not be generated by the closed-loop system. However, $L_a(\mathcal{S}^a/G)=K$ ensures the closed-loop system is always safe.
\end{remark}

Our goal is to control the system $G$ to be safe under cyber
attacks. The safety specification is described by a specification
language $K$ generated by a sub-automaton of $G$ as
\begin{align*}
	H = (Q_H, \Sigma, \delta_H, q_0),
\end{align*}
where $Q_H \subseteq Q$ and $\delta_H = \delta|_{Q_H \times \Sigma} \subseteq \delta$. Intuitively, $Q_H$ is the set of safe states and $Q-Q_H$ is the set of unsafe states.

To achieve this goal, a supervisor must be designed such that the large language of the supervised system $L_a(\mathcal{S}^a/G)$ is
equal to or contained in the specification language $K$. To design such a resilient supervisor, we first need to find the existence condition for a supervisor $\cS$ such that $L_a(\mathcal{S}^a/G)=K$. If the existence condition is not satisfied, we can than find a sublanguage of $K$ that satisfies the existence condition. Therefore, we want to solve the following control problem. 

\textbf{Safety Supervisory Control Problem for Discrete Event Systems
	under Cyber Attacks (SCPDES-CA):} Consider a discrete event system
$G$ under cyber attacks in the observation channel described by
$\Phi^\pi$, and in the control channel described by $\Delta$. For
a non-empty closed specification language $K \subseteq L(G)$
modeled as a sub-automaton $H \sqsubseteq G$, find a supervisor
$\cS$ such that $L_a(\cS^a/G) = K$.

To solve SCPDES-CA, we first investigate how to estimate states
and then consider state-estimate-based supervisory control.

\section{State estimation and state-estimate-based supervisors}\label{s4}

After observing a string $t \in \Phi ^\pi (L(G))$, the set of
possible states that $G$ may be in is called the state estimate
and is defined as
\begin{equation} \label{Equation3}
	\begin{split}
		SE^\pi_G(t)= \{& q \in Q: (\exists s \in L(G)) \\
		& t \in \Phi ^\pi (s) \wedge \delta  (q_0, s)=q \}.
	\end{split}
\end{equation}

\begin{remark} \rm
	Note that $\Phi ^\pi (s)$ is not prefix-closed. If $t \in
	\overline{\Phi ^\pi (s)} - \Phi ^\pi (s)$ is observed, it
	means that more events will be observed before any new event
	occurs in $G$. We can wait for more observations before
	making a control decision. Hence, there is no harm to let $SE^\pi_G(t)= \emptyset$ for $t \in \overline{\Phi ^\pi (s)} - \Phi ^\pi (s)$ as in the above definition.
\end{remark}

To obtain state estimate $SE^\pi_G(t)$, we do the following. For each transition $tr\in \delta^a$, let us assume that $A_{tr}$ is marked by an automaton $F_{tr}$. In other words, $A_{tr} = L_m (F_{tr})$ for some
$$
F_{tr} = (Q_{tr} , \Sigma, \delta_{tr}, q_{0,tr}, Q_{m,tr}).
$$

We replace each transition $tr=(q, \sigma, q') \in \delta ^a$ by
$(q, F_{tr}, q')$ as follows.
\begin{align}\label{equation of G replace}
	G_{tr \rightarrow (q, F_{tr}, q')} = ( Q \cup
	Q_{tr}, \Sigma , \delta_{tr \rightarrow (q,
		F_{tr}, q')}, q_0 )
\end{align}
where $\delta_{tr \rightarrow (q, F_{tr}, q')} = (\delta - \{ (q, \sigma, q') \} ) \cup \delta_{tr} \cup \{ (q, \varepsilon, q_{0,tr}) \} \cup \{ (q_{m,tr}, \varepsilon, q'): q_{m,tr} \in Q_{m,tr} \}$. 
In other words, $G_{tr \rightarrow (q, F_{tr}, q')}$ contains all
transitions in $G$ and $F_{tr}$, except $(q, \sigma, q')$, plus the
$\varepsilon$-transitions from $q$ to the initial state of
$F_{tr}$ and from marked states of $F_{tr}$ to $q'$.

Denote the automaton after replacing all transitions subject to
attacks as
\begin{equation}\label{equation of G diamond}
G^\diamond=( Q^\diamond, \Sigma , \delta^ \diamond , q_0, Q^\diamond_m)=( Q \cup \hat{Q},
\Sigma , \delta^ \diamond , q_0, Q)
\end{equation}
where $\hat{Q}$ is the set of states added during the replacement
and $Q^\diamond_m=Q$ is the set of marked states. Note that $G^\diamond$ is a nondeterministic automaton, that is, $\delta^ \diamond$ is a mapping $\delta^ \diamond : Q^\diamond \times \Sigma \rightarrow 2^{Q^\diamond}$.

To model the partial observation, we replace unobservable
transitions by $\varepsilon$-transitions and denote the resulting
automaton as
\begin{equation}\label{equation of G diamond varepsilon}
	G_{\varepsilon}^\diamond=( Q \cup \hat{Q}, \Sigma , \delta
	_{\varepsilon}^\diamond , q_0, Q)
\end{equation}
where $\delta _{\varepsilon}^\diamond = \{ (q, \sigma, q'): (q, \sigma,
q') \in \delta ^\diamond \wedge \sigma \in \Sigma _o \} \cup \{ (q,
\varepsilon, q'): (q, \sigma, q') \in \delta ^\diamond \wedge \sigma
\not\in \Sigma _o \} $. $G^\diamond_\varepsilon$ is also a
nondeterministic automaton.

$G_{\varepsilon}^\diamond$ can be used to calculate $\Phi ^\pi (L(G))$ as shown in the following theorem. Proofs of all the results are in the appendices. 

\begin{theorem}
	\label{theorem1} \rm
	The set of strings that can be observed under partial observation
	and cyber attacks in observation channel, $\Phi ^\pi (L(G))$, is
	given by
	$$
	\Phi ^\pi (L(G)) = L_m(G_{\varepsilon}^\diamond).
	$$	
\end{theorem}
where $L_m(G_{\varepsilon}^\diamond)= \{ s \in \Sigma_o^* : \delta _{\varepsilon}^\diamond (q_0, s) \cap Q \not= \emptyset \}$ is the language marked by $G_{\varepsilon}^\diamond$.

To obtain the state estimates, we convert $G_{\varepsilon}^\diamond$ to a
deterministic automaton $G_{obs}^\diamond$, called CA-observer, using
operator $OBS$ as follows.
\begin{equation}\label{equation of ca-obs}
	\begin{aligned}
		G_{obs}^\diamond & = OBS(G_{\varepsilon}^\diamond) =(X,\Sigma _o, \xi, x_0, X_m) \\
		&= Ac(2^{Q \cup \hat{Q}},\Sigma _o,
		\xi, UR(\{ q_0\}), X_m), 
	\end{aligned}
\end{equation}
where $Ac(.)$ denotes the accessible part; $UR(.)$ is the
unobservable reach defined, for $x \subseteq Q \cup \hat{Q}$, as
$$
UR(x) = \{ q \in Q \cup \hat{Q}: (\exists q' \in x) q\in \delta
_{\varepsilon}^\diamond (q', \varepsilon) \}.
$$

The transition function $\xi$ is defined, for $x \in X$ and
$\sigma \in \Sigma _o$ as
$$
\xi (x, \sigma) = UR(\{q \in Q \cup \hat{Q}: (\exists q' \in
x)q\in \delta _{\varepsilon}^\diamond (q',\sigma) \}).
$$
The marked states are defined as
$$
X_m = \{x \in X: x \cap Q \not = \emptyset \}.
$$
It is well-known that $L(G_{\varepsilon}^\diamond)=L(G_{obs}^\diamond)$ and
$L_m(G_{\varepsilon}^\diamond)=L_m(G_{obs}^\diamond)$. Hence, $\Phi ^\pi (L(G)) = L_m(G_{obs}^\diamond)$.

With observer $G_{obs}^\diamond$, we can calculate the estimate for every
observation in $\Phi ^\pi (L(G))$ as follows.
\begin{theorem} \rm
	\label{theorem2}
	After observing $t \in \Phi ^\pi (L(G))
	=L_m(G_{obs}^\diamond)$, the state estimate $SE^\pi_G(t)$ is given by
	$$
	SE^\pi_G(t)= \xi (x_0, t) \cap Q.
	$$
\end{theorem}

\begin{remark} \rm	
	Theorem \ref{theorem2} considers $t \in \Phi ^\pi
	(L(G))$ only. For $t \in \overline{\Phi ^\pi (L(G))} - \Phi ^\pi
	(L(G))$, we do not care what $SE_G^\pi (t)$ is. To be consistent,
	we can define $SE_G^\pi (t)$ as $SE_G^\pi (t) = \xi (x_0,t) \cap Q
	= \emptyset$ with no impact on control, because the control
	decision issued for $t \in \overline{\Phi ^\pi (L(G))} - \Phi ^\pi
	(L(G))$ does not change the dynamics of the closed-loop system.
\end{remark}

The procedure of calculating the stat estimates  under the observation $\Phi^\pi$ is presented by Algorithm \ref{Algorithm of state estimation}.
\begin{breakablealgorithm}\label{Algorithm of state estimation}
	\caption{Calculate State Estimates Under the Observation $\Phi^\pi$} 		
	\begin{algorithmic}[1]
		\Require $G, \mathbb{A}, t\in\Phi^\pi(L(G))$. 
		\Ensure $SE^\pi_G(t)$. 
		\State Construct automata $F_{tr} = (Q_{tr} , \Sigma, \delta_{tr}, q_{0,tr}, Q_{m,tr})$ as $A_{tr}(\in \mathbb{A})=L_m(F_{tr} )$.
		\State Construct  automata  $G^\diamond$ by replacing all $tr\in\sigma^a$ in $G$ with $F_{tr}$, shown as Equation(\ref{equation of G replace}) and (\ref{equation of G diamond}).
		\State Repalce all unobservable trasnsitons with $\varepsilon$ in $G^\diamond$ and construct $G_{\varepsilon}^\diamond$ as Equation(\ref{equation of G diamond varepsilon}).
		\State Construct CA-observer as Equation(\ref{equation of ca-obs}), the transition function of which is $\xi$.
		\State Output 	$SE^\pi_G(t)= \xi (x_0, t) \cap Q.$
	\end{algorithmic}
\end{breakablealgorithm}

Let us use an example to illustrate how to calculate state
estimates for observations.
\begin{example}\label{example obs} \rm
	Let us continue Example \ref{attackmodel} with discrete event
	system $G$ shown in Fig. \ref{attacks}. The two automata $F_{tr_1}$
	and $F_{tr_2}$ corresponding to $A_{tr_1}$ and $A_{tr_2}$ for
	$tr_1=(2,\lambda, 3)$ and $tr_2=(3, \mu, 1)$ are shown
	in Fig. \ref{Ft1 and Ft2}.
	\begin{figure}[htb]
		\centering
		\includegraphics[scale=1]{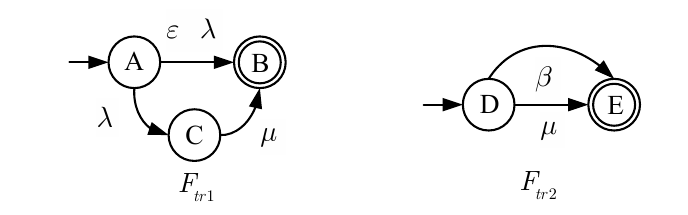}
		\caption{Automata $F_{tr_1}$ and $F_{tr_2}$ used
			to describe languages $A_{tr_1}$ and $A_{tr_2}$ }
		\label{Ft1 and Ft2}
	\end{figure}
	
	We replace transitions $tr_1$ and $tr_2$ with the corresponding
	automata $F_{tr_1}$ and  $F_{tr_2}$ to obtain $G^\diamond$, which is shown
	in Fig. \ref{Gr}. Since all events are observable,
	$G^\diamond_\varepsilon=G^\diamond$.
	\begin{figure}[htb]
		\centering
		\includegraphics[scale=1]{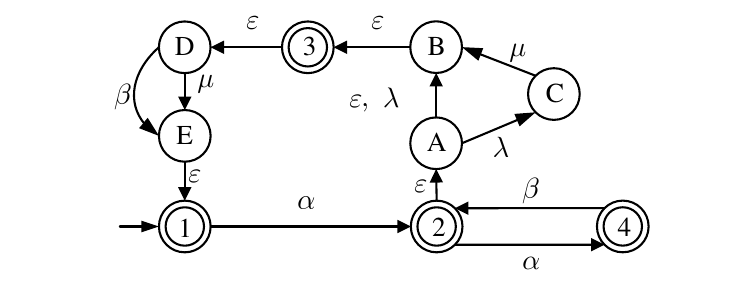}
		\caption{Automaton $G^\diamond$}
		\label{Gr}
	\end{figure}

	We then construct the observer $G^\diamond_{obs}$ for automaton
	$G^\diamond$, which is shown in Fig. \ref{Gobs}.
	\begin{figure}[htb]
		\centering
		\includegraphics[scale=1]{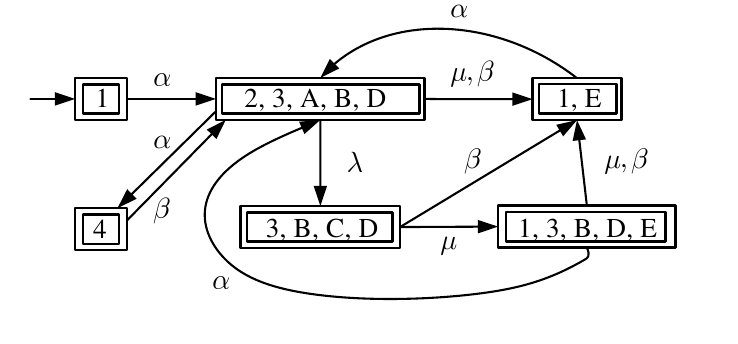}
		\caption{Observer $G^\diamond_{obs}$ for automaton $G^\diamond$}
		\label{Gobs}
	\end{figure}
	
	Using the observer $G^\diamond_{obs}$, we can calculate the state
	estimate for any observation $t \in \Phi ^\pi (L(G))$. For example,
	for $t=\alpha\lambda\mu$, we have
	\begin{equation}
		\xi(t)=\{1, 3, B, D, E\}. \notag
	\end{equation}
	Hence
	\begin{equation}
		SE^\pi_G(t)= \xi (x_0, t) \cap Q=\{1, 3, B, D, E\}\cap Q=\{1, 3\}. \notag
	\end{equation}
	
\end{example}

Using the method of calculating state estimates proposed above, we
can construct a state-estimate-based supervisor $\mathcal{S}_{CA}$, called
CA-supervisor, as follows.

For the specification automaton $H$, after observing a string $t \in
\Phi ^\pi(L(H))$, the state estimate of close-loop system is
$SE^\pi_H(t)$ and defined as\footnote{Note that the states that
	may be reached by the close-loop system are those in $H$, not in
	$G$. Hence we calculate the state estimates for $H$, not for $G$
	to obtain a supervisor.}

\begin{equation} \label{SEH}
	\begin{split}
		SE^\pi_H(t)=  \{ &q \in Q_H: (\exists s \in L(H)) \\
		& t \in \Phi ^\pi (s) \wedge \delta _H (q_0, s)=q \}.
	\end{split}
\end{equation}

To ensure that the supervised system will not enter unsafe states
in $Q-Q_H$, the set of events that need to be disabled is
\begin{equation} \label{gamma}
	\begin{split}
		\eta(SE^\pi_H(t))=\{ & \sigma\in \Sigma: (\exists q \in SE^\pi_H(t))
		\delta(q,\sigma)\in Q-Q_H\}.
	\end{split}
\end{equation}
Since a supervisor cannot disable uncontrollable events,
$\cS_{CA}(t)$ is defined as
\begin{equation} \label{Equation6}
	\cS_{CA}(t)=\begin{cases}
		(\Sigma -\eta ( SE^\pi_H(t)))\cup \Sigma _{uc}&\text{if $ t\in \Phi ^{\pi}(L(H))$}\\
		\Sigma _{uc}&\text{others}
	\end{cases}
\end{equation}

\begin{remark}
	Constructing $\cS_{CA}$ needs to calculate the state estimate for every observable string which can be implemented online with polynomial complexity or offline with exponential complexity since we need to construct the observer for the specification automaton $H$ and sensor attack automata.
\end{remark}

In the next section, we show that $\cS_{CA}$ defined above is
indeed a solution to the supervisory control problem under cyber
attacks when some necessary and sufficient conditions are
satisfied.

\section{Solutions to the supervisory control problem under cyber attacks}\label{s5}

Let us now solve SCPDES-CA. We first extend controllability and
observability to capture the impacts of cyber attacks on the
control channel and observation channel.

We call the extended controllability CA-controllability, which is
defined as follows.

\begin{definition} \label{Definition3} \rm
	A closed language $K \subseteq L(G)$ is {\em CA-controllable} with
	respect to $L(G)$, $\Sigma _{uc}$, and $\Sigma ^a_c$ if
	$$ \label{C}
	K (\Sigma _{uc} \cup \Sigma ^a_c) \cap L(G) \subseteq K. \notag
	$$
\end{definition}

In other words, $K$ is CA-controllable with respect to $L(G)$,
$\Sigma _{uc}$ and $\Sigma ^a_c$ if and only if $K$ is
controllable with respect to $L(G)$ and $\Sigma _{uc} \cup \Sigma
^a_c$. Intuitively, this is because attackers can enable events in
$\Sigma ^a_c$ and hence makes them uncontrollable as far as the
large language is concerned.

We call the extended observability CA-observability, which is
defined as follows.

\begin{definition} \label{Definition4} \rm
	A closed language $K \subseteq L(G)$ is {\em CA-observable} with
	respect to $L(G)$, $\Sigma _{o}$, $\Sigma ^a_o$ and $\Phi^\pi$ if
	\begin{equation} \label{Equation7}
		\begin{split}
			& (\forall s \in \Sigma ^*)(\forall \sigma \in \Sigma) (s \sigma
			\in K \\
			& \Rightarrow (\exists t \in \Phi ^\pi (s))(\forall s' \in (\Phi
			^\pi)^{-1} (t)) \\
			& (s' \in K \wedge s' \sigma \in L(G) \Rightarrow s' \sigma \in
			K)),
		\end{split}
	\end{equation}
	where $(\Phi ^\pi)^{-1}$ is the inverse mapping of $\Phi ^\pi$: for $t
	\in \Phi ^\pi (L(G))$,
	$$
	(\Phi ^\pi)^{-1} (t) = \{ s' \in L(G) : t \in \Phi ^\pi (s') \}.
	$$
	
\end{definition}

With  CA-controllability and CA-observability defined, we have the
following theorem.

\begin{theorem} \label{theoreme3} \rm
	Consider a discrete event system $G$ under cyber attacks. For a
	nonempty closed language $K \subseteq L(G)$, SCPDES-CA is
	solvable, that is, there exists a supervisor $\cS$ such that
	$L_a(\cS^a/G)=K$ if and only if $K$ is CA-controllable with
	respect to $L(G)$, $\Sigma _{uc}$, and $\Sigma ^a_c$ and
	CA-observable with respect to $L(G)$, $\Sigma _{o}$, $\Sigma
	^a_o$ and $\Phi^\pi$. Furthermore, if SCPDES-CA is solvable, then CA-supervisor
	$S_{CA}$ defined in Equation (\ref{Equation6}) is a valid
	solution, that is,		
	$$
	L_a(\cS^a_{CA}/G)=K.
	$$
\end{theorem}

Let us use an example to illustrate these results.

\begin{example}\label{E3} \rm
	Let us again consider the discrete event system given in Example
	\ref{attackmodel}. The specification language $K$ is described by
	a sub-automaton $H$ shown in Fig. \ref{K}. Consider the following
	two cases.
	\begin{figure}[htb]
		\centering
		\includegraphics[scale=1]{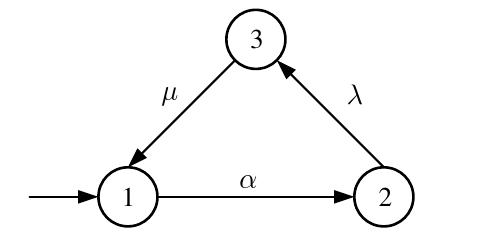}
		\caption{Automaton $H$ for the specification language $K$}
		\label{K}
	\end{figure}
	
	{\em Case 1}: $\Sigma_o=\Sigma_c=\Sigma$, $\Sigma^a_o=\{\lambda,
	\mu\}$ and $\Sigma^a_c=\{\alpha, \beta\}$. In this case, $K$ is
	not CA-controllable because
	\begin{align*}
		& (\exists s=\alpha \in \Sigma ^*) (\exists \sigma = \alpha \in
		\Sigma) \\
		& s \in K \wedge \sigma \in \Sigma _{uc} \cup \Sigma ^a_c \wedge s
		\sigma \in L(G) \wedge s \sigma \not\in K.
	\end{align*}
	By Theorem 3, no supervisor exists.
	
	{\em Case 2}: $\Sigma_o=\Sigma_c=\Sigma$, $\Sigma^a_o=\{\lambda,
	\mu\}$ and $\Sigma^a_c=\{ \beta\}$. In this case, $K$ is
	CA-controllable because
	\begin{align*}
		& (\forall s \in \Sigma ^*) (\forall \sigma \in \Sigma)  s \in K
		\wedge s \sigma \in L(G) \wedge s \sigma \not\in K \\
		& \Rightarrow \sigma = \alpha \not\in \Sigma _{uc} \cup \Sigma
		^a_c \\
		\Leftrightarrow & (\forall s \in \Sigma ^*) (\forall \sigma \in
		\Sigma) s \in K \wedge \sigma \in \Sigma
		_{uc} \cup \Sigma ^a_c \wedge s \sigma \in L(G)  \\
		& \Rightarrow s \sigma \in K .
	\end{align*}
	Let us show that $K$ is also CA-observable, that is, Equation
	(\ref{Equation7}) is satisfied.
	Clearly, $s \sigma \in K$ if and only if (1) $s \sigma =
	(\alpha\lambda\mu)^n \alpha$, or (2) $s \sigma =
	(\alpha\lambda\mu)^n \alpha \lambda$, or (3) $s \sigma =
	(\alpha\lambda\mu)^n$, for $n=1,2,...$.
	
	For (1) $s \sigma = (\alpha\lambda\mu)^n \alpha$, we have
	\begin{align*}
		& s \sigma = (\alpha\lambda\mu)^n \alpha \\
		\Rightarrow & s = (\alpha\lambda\mu)^n \wedge \sigma = \alpha \\
		\Rightarrow & s = (\alpha\lambda\mu)^n \wedge \sigma = \alpha
		\wedge (\exists t = (\alpha\lambda\mu\mu)^n \in \Phi ^ \pi (s)) \\
		& (\forall s' = (\alpha\lambda\mu)^n \in (\Phi ^ \pi)^{-1} (t))
		(s' = (\alpha\lambda\mu)^n \in K \\
		& \wedge s' \sigma = (\alpha\lambda\mu)^n \alpha \in L(G)
		\Rightarrow s' \sigma = (\alpha\lambda\mu)^n \alpha \in K).
	\end{align*}
	Hence, Equation (\ref{Equation7}) is satisfied. Similarly, we can
	show that, for (2) $s \sigma = (\alpha\lambda\mu)^n \alpha
	\lambda$, and (3) $s \sigma = (\alpha\lambda\mu)^n$, Equation
	(\ref{Equation7}) is also satisfied. Therefore, $K$ is
	CA-observable.
	
	By Theorem 3, CA-supervisor $\cS_{CA}$ defined in Equation
	(\ref{Equation6}) is a valid supervisor such that
	$$
	L_a(\cS^a_{CA}/G)=K.
	$$
	The observer $H^\diamond_{obs}$ can be calculated in the same way as
	$G_{obs}^\diamond$ and is shown in Fig. \ref{Hobs}.
	\begin{figure}[htb]
		\centering
		\includegraphics[scale=1]{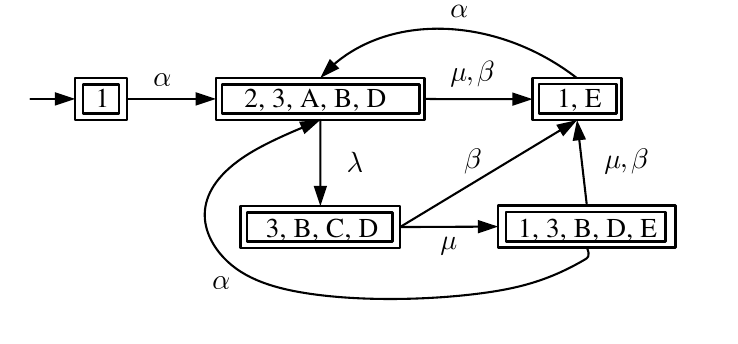}
		\caption{Observer $H^\diamond_{obs}$}
		\label{Hobs}
	\end{figure}
	
	From $H$, we know when the system enters state $2$, event $\alpha$
	should be disabled. Therefore, for any observable string $t$ which
	drives $H^\diamond_{obs}$ to state $\{2,3,A,B,D\}$, let
	$$
	\cS_{CA}(t)=\{\lambda, \mu, \beta\}.
	$$
	For the other observable strings $t$, let
	$$
	\cS_{CA}(t)=\{\alpha, \lambda, \mu, \beta\}.
	$$
\end{example}
\section{Maximally-permissive supervisor}\label{s6}

Unlike the traditional supervisory control problem, the supervisor that solves SCPDES-CA is not unique. 

Note that if a supervisor $\cS$ is a solution to SCPDES-CA, then the large language $L_a(\cS^a/G)$ equals to $K$ which is generated by $H$. Hence, the supervisor $\cS$ will never observe  any string $t\in\Sigma_o^* -\Phi^\pi(L(H))$.
Therefore, the control command $\cS(t)$ for $t\in\Sigma_o^*
-\Phi^\pi(L(H))$ is inconsequential in the sense that it will not
alter the large language. Hence, we only need to consider
supervisors defined as $\cS : \Phi ^\pi ( L(H)) \rightarrow
\Gamma$.

Given a finite or infinite set of supervisors $\{ \cS_j : j=1,2,3,
... \}$ ($j=1,2,3,...,J$ if the set is finite), their union,
denoted as $\cup \cS_{j}$, is defined as
\begin{equation} \label{solutioncup}
	(\forall t\in \Phi^\pi(L(H))) (\cup \cS_{j})(t)= \cS_1 (t) \cup \cS_2
	(t) \cup \cS_3 (t) \cup  ...
\end{equation}

From the above definition, it is clear that
\begin{equation}
	(\forall j) L_a(\cS_j^a/G) \subseteq L_a(\cup \cS^a_{ j}/G).
	\label{inclusion}
\end{equation}

Denote the set of all supervisors that solve SCPDES-CA as
\begin{equation}
	\Omega=\{\cS:L_a(\cS^a/G)=K\} \label{solution}
\end{equation}

The following proposition says that the set $\Omega$ is closed
under arbitrary union.

\begin{proposition} \label{proposition 1} \rm
	If supervisors $\{ \cS_j, j=1,2,3, ... \}$ are solutions to 
	SCPDES-CA, then their union is also a solution to SCPDES-CA,
	that is, for any set of supervisors $\{ \cS_j, j=1,2,3, ... \}$,
	\begin{equation}
		((\forall j) \cS_j\in \Omega) \Rightarrow \cup \cS_{j} \in \Omega. \label{proposition1}
	\end{equation}	
\end{proposition}

Given supervisors $\cS_1$ and $\cS_2$, we say $\cS_1$ is less permissive than $\cS_2$, denoted as $\cS_1 \leq \cS_2$, if $(\forall t\in \Phi^\pi(L(H))) \cS_1(t) \subseteq \cS_2(t)$. 
	We say $\cS_1$ is strictly less permissive than $\cS_2$, denoted as $\cS_1 \textless \cS_2$, if $\cS_1 \leq \cS_2 \wedge (\exists t\in \Phi^\pi(L(H))\cS_1(t) \subset \cS_2(t))$.

By Proposition \ref{proposition 1}, we conclude that there exists a unique supervisor in $\Omega$, denoted by $\cS^\circ$, that is more permissive then any other supervisor in $\Omega$, that is,
	$$
	(\forall \cS_j \in \Omega)\cS_j \leq \cS^\circ.
	$$

$\cS^\circ$ is called the maximally-permissive supervisor. It can be obtained as follows. Enumerate all
	supervisors in $\Omega$ as
	$$
	\Omega = \{ \cS_j, j=1,2,3, ... \}.
	$$
	Then
	$$
	\cS^\circ = \cup \cS_{j}.
	$$	
By Proposition \ref{proposition 1}, $\cS^\circ = \cup \cS_{j} \in \Omega$ is well defined
and unique. Furthermore, by Equation (\ref{inclusion}),
$$
(\forall \cS_{j} \in \Omega) L_a(\cS_j^a/G) \subseteq
L_a(\cup \cS^a_{ j}/G).
$$

Calculating the maximally-permissive supervisor that solves 
SCPDES-CA using $\cS^\circ = \cup \cS_{j}$ is not practical
because it requires to find all $\cS_{j} \in \Omega$. This is
however unnecessary because the following theorem says that the
CA-supervisor $\cS_{CA}$ defined in the previous section is the maximally-permissive supervisor.

\begin{theorem} \label{Theorem 4}\rm
	Assume that SCPDES-CA has a solution, that is,  $\Omega \ne
	\emptyset$. Then $\cS_{CA}$ given in Equation (\ref{Equation6}) is
	the maximally-permissive supervisor that solves SCPDES-CA, that is,
	$$\cS_{CA}=\cS^\circ.$$
\end{theorem}

\begin{example} \rm
	Let us continue Example \ref{E3}. We assume
	$\Sigma^a_c=\{\beta\}$. From Example \ref{E3}, we know SCPDES-CA
	has solutions. Hence the supervisor $\cS_{CA}$ is a solution.
	
	In fact, we can find another valid supervisor $\cS_1$ which is
	defined as follows.
	
	For any observable string $t$ which drives $H^\diamond_{obs}$ to state
	$\{2,3,A,B,D\}$, let
	$
	\cS_1(t)=\{\lambda, \mu, \beta\}
	$.
	
	For any observable string $t$ which drives $H^\diamond_{obs}$ to state
	$\{3 ,B, C, D\}$, let
	$
	\cS_{1}(t)=\{\beta, \mu\}
	$.
	
	For the other observable strings $t$, let
	$
	\cS_{1}(t)=\{\alpha , \mu, \beta \}
	$.
	
	It is not difficult to verify that $L_a((\cS_1)^a/G) = K$ and $\cS_1<\cS_{CA}$.
	
\end{example}

\section{Observation-based sensor attacks}\label{s7}

In the previous sections, we have investigated transition-based
sensor attack strategies. In practice, however, an attacker can
make different sensor attack decisions for different observations.
Hence, in this section, we extend our results to this general case
by considering observation-based sensor attack strategies of the
form
$$
\varpi: P(L(G)) \times \Sigma_o^a \rightarrow 2^{\Sigma_o^*}
$$
That is, after an observable string $t$ is observed, the attacker
changes an attacked event $\sigma\in \Sigma^a_o$ into a string
$t'$ in $\varpi(t,\sigma)$ which is a set of observable strings (a
language). The supervisor will see $t'$, instead of $\sigma$. In
other words, we extend sensor attack strategies from $\pi (tr)$ to
$\varpi(t,\sigma)$.

Since the number of observable strings is infinite, if a sensor
attack strategy can be implemented by an attacker finitely, it
must be described by a structure with finite states which models
all possible situations for which different attack decisions are
assigned. Without loss of generality, we assume such a structure
is an automaton defined as
$$
SA=(Z,\Sigma_o,\delta_{SA}, z_0)
$$
where $Z$ is a finite set of states and $\delta_{SA}: Z \times
\Sigma_o \rightarrow Z$ is the transition function. We require
$P(L(G)) \subseteq L(SA)$.

A sensor attack strategy is then defined on the state set $Z$ as
$$
\omega: Z \times \Sigma_o^a \rightarrow 2^{\Sigma_o^*}.
$$
Since the sensor attack strategy does not change the observations
of events in $\Sigma_o-\Sigma_o^a$, for any state $z\in Z$ and
event $\sigma\in \Sigma_o-\Sigma_o^a$, we let $\omega(z,
\sigma)=\{\sigma\}$. We also let $\omega(z, \varepsilon)
=\{\varepsilon\}$. From now on, we will consider the sensor attack
strategy $\omega$ instead of $\varpi$.

Under the sensor attack strategy $\omega$, the possible
observations for an observable string $t\in P(L(G))$, denoted as
$\Phi^\omega(t)$, can be calculated recursively as
\begin{align*}
	& \Phi^\omega(\varepsilon)=\{\varepsilon\} \\
	& \Phi^\omega(t\sigma)= \Phi^\omega(t)\omega(\delta_{SA}(z_0, t),\sigma).
\end{align*}

After the occurrence of $s\in L(G)$, a supervisor $\cS$ under
observation-based sensor attack observes one of the string $t \in
\Phi^\omega (P(s))$.  Hence, $S$ is a mapping
$$
\mathcal{S} : \Phi^\omega (P(L(G))) \rightarrow 2^{\Sigma}.
$$

The behavior of supervised system under observation-based sensor
attack is also described by the large language $L_a(\cS/G)$, whose definition is similar to
Definitions \ref{Definition1}with $\Phi
^\pi (s)$ replaced by $\Phi^\omega (P(s))$.

The control problem we want to solve in this section is as
follows.

\textbf{Supervisory Control Problem for Discrete Event Systems
	under Observation-Based Cyber Attacks (SCPDES-OBCA):} Consider a
discrete event system $G$ under cyber attacks in the observation
channel described by $\Phi^\omega$, and in the control channel
described by $\Delta$. For a non-empty closed specification
language $K \subseteq L(G)$ modeled as a sub-automaton $H
\sqsubseteq G$, find a supervisor $\cS$ such that $L_a(\cS^a/G) =
K$.

The approach for solving SCPDES-OBCA is similar to that for
SCPDES-CA. We first need to calculate state estimates under
observation-based sensor attack. For an observation $t \in
\Phi^\omega (P(L(G)))$, the state estimate is defined as
\begin{equation} \label{SEGt}
	\begin{split}
		SE_G^\omega(t)=\{&q\in Q: (\exists s \in L(G)) t \in \Phi ^\omega (P(s))\wedge\\
		& \delta(q_0 ,s)=q\}.
	\end{split}
\end{equation}

To calculate $SE_G^\omega(t)$, we first convert the problem of
finding state estimates under observation-based sensor attack to
the problem of finding state estimates under transition-based
sensor attack. We then use the results of Section \ref{s4} to find
state estimates $SE_G^\omega(t)$ under observation-based sensor
attack. The following procedure can be used for the conversion.

Step 1. Take the parallel composition of $G$ and SA to obtain an
augmented automaton as
$$
\tilde{G}=G||SA=(Y, \Sigma, \tilde{\delta}, y_0)=Ac(Q\times Z,
\Sigma, \tilde{\delta}, (q_0, z_0)) .
$$

Step 2. Define $\tilde{\delta} ^a$ as
$$
\tilde{\delta} ^a = \{ (y_i, \sigma, y_j) \in \tilde{\delta} :
\sigma \in \Sigma _o^a \}.
$$

Step 3. For a transition $\tilde{tr} = (y_i, \sigma, y_j) =((q_i,
z_i), \sigma, $ $ (q_j, z_j)) \in \tilde{\delta} ^a$, let
$$
A_{\tilde{tr}} = \omega(z_i, \sigma).
$$

Step 4. Define a transition-based sensor attack strategy
$\tilde{\pi}$ as follows. For a transition $\tilde{tr} = (y_i,
\sigma, y_j) \in \tilde{\delta} ^a$,
$$
\tilde{\pi}(\tilde{tr})= A_{\tilde{tr}}.
$$

Note that since $P(L(G)) \subseteq L(SA)$, $\tilde{G}$ and $G$
generate the same language: $L(\tilde{G})=L(G)$.

Since $\tilde{\pi}$ is transition-based, we can use the results in
Section \ref{s2} to calculate $\tilde{\Theta}^{\tilde{\pi}}(s)$
and $\tilde{\Phi}^{\tilde{\pi}}(s)$ for any string $s\in
L(\tilde{G})$ as follows. 

Let
$s=\sigma_1\sigma_2\cdots\sigma_{|s|} \in L(\tilde{G})$ and $y_k =
\tilde{\delta} (y_0,\sigma _1\cdots\sigma _{k})$, $k= 1, 2, ..., |s|$. We
have
$$
\tilde{\Theta}^{\tilde{\pi}}(s) = L_1 L_2 ... L_{|s|},
$$
where
\begin{align*}
	L_k = \left\{ \begin{array}{ll} \{ \sigma_k \} & \mbox{if }
		\tilde{tr}=(y_{k-1}, \sigma _k, y_k)\not\in \tilde{\delta}^a \\
		A_{\tilde{tr}},  & \mbox{if } \tilde{tr}=(y_{k-1}, \sigma _k, y_k)
		\in \tilde{\delta}^a \end{array} \right .
\end{align*}
and
\begin{equation} \label{tPhi}
	\tilde{\Phi}^{\tilde{\pi}}(s)=P(\tilde{\Theta}^{\tilde{\pi}}(s)) =
	P(L_1)P(L_2) ... P(L_{k}).
\end{equation}

The following proposition shows the equivalence of the
observation-based sensor attack $\omega$ and the transition-based
sensor attack $\tilde{\pi}$.

\begin{proposition} \label{proposition 2} \rm
	Sensor attack strategies $\tilde{\pi}$ and $\omega$ result in the
	same observations, that is,
	$$
	(\forall s\in L(G))\tilde{\Phi}^{\tilde{\pi}}(s)=\Phi^\omega(P(s)).
	$$
\end{proposition}

Proposition \ref{proposition 2} allows us to convert the problem
of finding state estimates under observation-based sensor attack
$\omega$ to the problem of finding state estimates under
transition-based sensor attack $\tilde{\pi}$, which is defined as
\begin{equation} \label{SEt}
	\begin{split}
		SE^{\tilde{\pi}}_{\tilde{G}}(t)= \{& y=(q,z) \in Y: (\exists s \in
		L(\tilde{G})) \\
		& t \in \tilde{\Phi} ^{\tilde{\pi}} (s) \wedge \tilde{\delta}
		(y_0, s)=y \}.
	\end{split}
\end{equation}
$SE^{\tilde{\pi}}_{\tilde{G}}(t)$ can be calculated using the
methods in Section \ref{s4}. After $SE^{\tilde{\pi}}
_{\tilde{G}}(t)$ is calculated, the corresponding state estimate
in $G$ is given by
\begin{equation} \label{RSEt}
	\begin{split}
		R(SE^{\tilde{\pi}}_{\tilde{G}}(t))=\{q \in Q : (\exists z \in Z)
		(q,z) \in SE^{\tilde{\pi}}_{\tilde{G}}(t)\}.
	\end{split}
\end{equation}

The following theorem says that $R(SE^{\tilde{\pi}} _{\tilde{G}}
(t))$ is the desired state estimate.

\begin{theorem}\label{theorem 5} \rm
	For any observation $t \in \Phi^\omega (P(L(G)))$, its state
	estimate in $G$ is given by
	$$
	SE^\omega_G(t)=R(SE^{\tilde{\pi}}_{\tilde{G}}(t)).
	$$
\end{theorem}

Using the state estimate method proposed above, we derive a
state-estimate-based supervisor as follows. (1) Construct the
parallel composition $\tilde{H}=H||SA$. Since $H$ is a
sub-automaton of $G$, $\tilde{H}$ is a sub-automaton of
$\tilde{G}$. (2) Calculate state estimates $SE^{\tilde{\pi}}
_{\tilde{H}}(t)$ and $SE^\omega_H(t) =R(SE^{\tilde{\pi}}
_{\tilde{H}}(t))$ in the same way as $SE^{\tilde{\pi}}
_{\tilde{G}}(t)$ and $SE^\omega_G(t) =R(SE^{\tilde{\pi}}
_{\tilde{G}}(t))$ (replacing $G$ by $H$). (3) The
state-estimate-based supervisor $\tilde{\cS}_{CA}$ is given by
\begin{equation} \label{observation supervisor}
	\tilde{\cS}_{CA}(t)=\begin{cases}
		(\Sigma -\eta (SE_H^\omega(t))\cup \Sigma _{uc}&
		\text{if $t\in \Phi ^\omega(L(H))$}\\
		\Sigma _{uc}&\text{others}
	\end{cases}
\end{equation}
where
$$
\eta(SE^\pi_H(t))=\{ \sigma\in \Sigma: (\exists q \in
SE^\pi_H(t)) \delta(q,\sigma)\in Q-Q_H\}.
$$

We can now state the following necessary and sufficient condition
for solving SCPDES-OBCA.

\begin{theorem} \label{theoreme6} \rm
	Consider a discrete event system $G$ under cyber attacks in the
	observation channel described by $\Phi^\omega$, and in the control
	channel described by $\Delta$. For a non-empty closed
	specification language $K \subseteq L(G)$ modeled as a
	sub-automaton $H \sqsubseteq G$, SCPDES-OBCA is solvable, that is,
	there exists a supervisor $\cS$ such that $L_a(\cS^a/G)=K$ if and
	only if $K$ is CA-controllable with respect to $L(G)$, $\Sigma
	_{uc}$, and $\Sigma ^a_c$ and CA-observable with respect to
	$L(G)$, $\Sigma _{o}$, $\Sigma ^a_o$ and $\Phi^\omega$.
	Furthermore, if SCPDES-OBCA is solvable, then CA-supervisor
	$\tilde{\cS}_{CA}$ defined in Equation (\ref{observation
		supervisor}) is a valid solution, that is,
	$$
	L_a(\tilde{\cS}^a_{CA}/G)=K.
	$$
\end{theorem}

Note that in Theorem \ref{theoreme6}, CA-controllability is
identical to CA-controllability in Theorem \ref{theoreme3}, while
CA-observability is with respect to $\Phi^\omega$ rather than
$\Phi^\pi$.

\begin{example} \rm
	Let us again consider the discrete event system $G$ given in
	Example \ref{G} and the control specification automaton $H$ in
	Example \ref{E3}. We assume all events are controllable and observable;
	and $\Sigma_o^a=\{\lambda, \mu\}$ and $\Sigma_c^a=\{\beta\}$.	
	\begin{figure}[htb]
		\centering
		\includegraphics[scale=1]{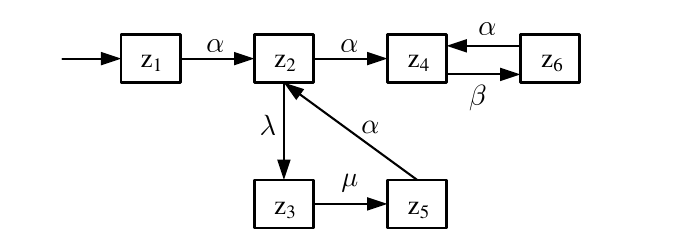}
		\caption{Automaton $SA$}
		\label{SA}
	\end{figure}
	
	The observation-based sensor attack strategy is given by $SA$  shown in Fig. \ref{SA} and the sensor attack
	strategy $\omega$ is defined as
	\begin{align*}
		\omega(z_2, \lambda)&=\{\varepsilon, \lambda, \lambda\mu\}\\
		\omega(z_3, \mu)&=\{\mu, \beta\}.
	\end{align*}
	
	We construct $\tilde{H}$  as shown in Fig. \ref{GSA}.
	\begin{figure}[htb]
		\centering
		\includegraphics[scale=1]{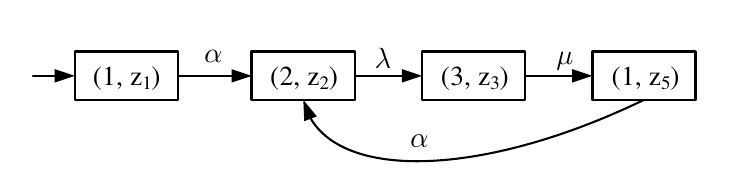}
		\caption{Automaton $\tilde{H}$}
		\label{GSA}
	\end{figure}
	
	With $\omega$, we define the transition-based sensor attack
	strategy $\tilde{\pi}$ as follows.
	\begin{align*}
		\tilde{\pi}((2, z_2), \lambda, (3, z_3))&= \omega( z_2, \lambda)\\
		\tilde{\pi}((3, z_3), \mu, (1, z_5))&= \omega( z_3, \mu).
	\end{align*}
	
	Under transition-based sensor attack strategy $\tilde{\pi}$, we
	calculate $(\tilde{H})^\diamond$ and  $(\tilde{H})^\diamond_{obs}$ as shown in
	Fig. \ref{GSAr} and Fig. \ref{GSAobs}, respectively.
	\begin{figure}[htb]
		\centering
		\includegraphics[scale=1]{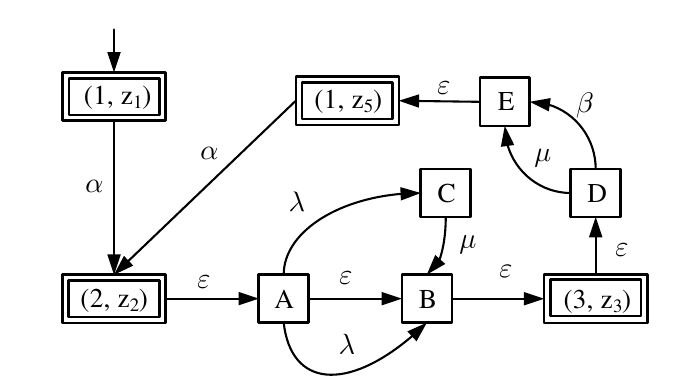}
		\caption{Automaton $(\tilde{H})^\diamond$}
		\label{GSAr}
	\end{figure}
	
	\begin{figure}[htb]
		\centering
		\includegraphics[scale=1]{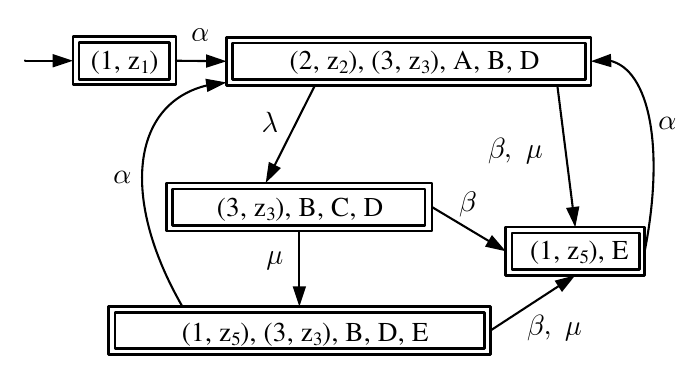}
		\caption{Observer $(\tilde{H})^\diamond_{obs}$ for automaton $(\tilde{H})^\diamond$}
		\label{GSAobs}
	\end{figure}
	
	Based on $(\tilde{H})^\diamond_{obs}$, we can obtain the CA-supervisor
	$\cS_{CA}$. For example, we can calculate the state estimate for
	observation $t=\alpha$ as
	\begin{align*}
		SE_H^\omega(\alpha)=&R(SE^{\tilde{\pi}}_{\tilde{H}}(\alpha))\\
		=&R((2,z_2),(3,z_3))=\{2,3\}
	\end{align*}
	and then the control as
	\begin{align*}
		\tilde{\cS}_{CA}(\alpha)=\{\beta,\lambda,\mu\}.
	\end{align*}
\end{example}

\section{Conclusions}\label{s8}

In this paper, we propose a method to control a given discrete event system under joint sensor and actuator attacks to be safe. By defining the upper-bound language and extending controllability and observability into CA-controllability and CA-observability, we successfully solve the supervisory control problem. In the future, we will consider how to control the system to achieve a maximal upper-bound sub-language when CA-controllability and/or CA-observability are not satisfied. We also want to find an effective method to check CA-observability. 

\appendices

\section{PROOF OF THEOREM \ref{theorem1}}

Since $G^\diamond$ is obtained by replacing all transitions subject to
attacks $(q, \sigma, q') \in \delta ^a$ by $(q, G_{(q, \sigma,
	q')}, q')$, we have
\begin{align*}
	& s \in L_m(G^\diamond) \\
	\Leftrightarrow & (\exists s'= \sigma _1 \sigma _2 ..., \sigma
	_{|s'|} \in L(G)) s \in L_1 L_2 ... L_{|s'|},
\end{align*}

where, for $k= 1, 2, ..., |s'|$
\begin{align*}
	L_k = \left\{ \begin{array}{ll}
		\{ \sigma _k \} & \mbox{if } (q_{k-1}, \sigma _k, q_k) \not\in \delta ^a \\
		L_m(G_{(q_{k-1}, \sigma _k, q_k)}) & \mbox{if } (q_{k-1}, \sigma
		_k, q_k) \in \delta ^a \end{array} \right.
\end{align*}
and $q_k = \delta(q_0,\sigma _1\cdots\sigma _{k})$.

Because $A_{(q_{k-1}, \sigma _k, q_k)}$ $= L_m (G_{(q_{k-1},
	\sigma _k, q_k)})$, we have
\begin{equation} \label{Equation2}
	\begin{split}
		L_k = \left\{ \begin{array}{ll}
			\{ \sigma _k \} & \mbox{if } (q_{k-1}, \sigma _k, q_k) \not\in \delta ^a \\
			A_{(q_{k-1}, \sigma _k, q_k)} & \mbox{if } (q_{k-1}, \sigma _k,
			q_k) \in \delta ^a \end{array} \right .
	\end{split}
\end{equation}

Obviously, Equations (\ref{Equation1}) and (\ref{Equation2}) are
identical. Thus,
\begin{align*}
	& s \in L_m(G^\diamond) \\
	\Leftrightarrow & (\exists s'= \sigma _1 \sigma _2 ..., \sigma
	_{|s'|} \in L(G)) s \in L_1 L_2 ... L_{|s'|} \\
	\Leftrightarrow & (\exists s'= \sigma _1 \sigma _2 ..., \sigma
	_{|s'|} \in L(G)) s \in \Theta ^\pi(s') \\
	\Leftrightarrow & s \in \Theta^\pi(L(G)).
\end{align*}
In other words,
\begin{equation} \label{Equation4}
	\begin{split}
		\Theta^\pi(L(G))=L_m(G^\diamond).
	\end{split}
\end{equation}

Because $G^\diamond_\varepsilon$ is obtained from $G^\diamond$ by replacing
unobservable events with the empty string $\varepsilon$, we have
\begin{align*}
	L_m(G^\diamond_\varepsilon)=P(L_m(G^\diamond)).
\end{align*}

Therefore,
\begin{align*}
	\Phi^\pi(L(G))=P(\Theta^\pi(L(G)))=P(L_m(G^\diamond))=L_m(G^\diamond_\varepsilon).
\end{align*}

\section{PROOF OF THEOREM \ref{theorem2}}
Since $G_{obs}^\diamond$ is the observer of $G_{\varepsilon}^\diamond$, by the
property of observer (see, for example, Proposition 3.1 of
\cite{lin2014control}), we have
\begin{align*}
	& q \in \xi (x_0, t) \cap Q \\
	\Leftrightarrow & (\exists s \in L(G^\diamond)) q \in \delta^\diamond (q_0, s)
	\wedge t = P (s) \wedge q \in Q \\
	\Leftrightarrow & (\exists s \in {L_m(G^\diamond)}) q \in\delta^\diamond  (q_0, s)
	\wedge t = P (s) \\
	& (\mbox{since $Q$ is the set of marked states of $G^\diamond$}) \\
	\Leftrightarrow & (\exists s \in {\Theta^\pi(L(G))})q \in \delta^\diamond  (q_0,
	s) \wedge t = P (s) \\
	& (\mbox{by Equation (\ref{Equation4})}) \\
	\Leftrightarrow & (\exists s' \in {L(G)}) s \in \Theta ^\pi(s')
	\wedge q \in \delta^\diamond (q_0, s) \wedge t = P (s) \\
	& (\mbox{by Equation (\ref{Equation5})}) \\
	\Leftrightarrow & (\exists s' \in {L(G)}) s \in \Theta ^\pi(s')
	\wedge \delta (q_0, s')=q \wedge t = P (s) \\
	& (\mbox{by the construction of $G^\diamond$, } q \in\delta^\diamond (q_0, s)\\
	&\Leftrightarrow \delta (q_0, s')=q) \\
	\Leftrightarrow & (\exists s' \in {L(G)}) \delta (q_0, s')=q
	\wedge t \in P (\Theta ^\pi(s')) \\
	\Leftrightarrow & (\exists s' \in {L(G)}) \delta (q_0, s')=q
	\wedge t \in \Phi ^\pi(s') \\
	\Leftrightarrow & q \in  SE^\pi_G(t) \\
	& (\mbox{by Equation (\ref{Equation3})})
\end{align*}

\section{PROOF OF THEOREM \ref{theoreme3}}
We first show that if $K$ is CA-observable with respect to $L(G)$,
$\Sigma _{o}$, $\Sigma ^a_o$ and $\Phi^\pi$, then, for all $s \in
\Sigma ^*$ and $\sigma \in \Sigma$,
\begin{equation} \label{Equation8}
	\begin{split}
		& s \sigma \in K \\
		\Leftrightarrow & s \in K \wedge s \sigma \in L(G) \\
		& \wedge (\exists t \in \Phi ^\pi(s)) (\forall s' \in
		(\Phi ^\pi)^{-1} (t)) \\
		&(s' \in K \wedge s' \sigma \in L(G) \Rightarrow s' \sigma \in K).
	\end{split}
\end{equation}

($\Rightarrow$) By the definition of CA-observability, Equation
(\ref{Equation7}), we have
\begin{align*}
	& s \sigma \in K \\
	\Rightarrow & (\exists t \in \Phi ^\pi (s))(\forall s' \in (\Phi
	^\pi)^{-1} (t)) \\
	& (s' \in K \wedge s' \sigma \in L(G) \Rightarrow s' \sigma \in K)
	\\
	\Rightarrow & s \in K \wedge s \sigma \in L(G) \\
	& \wedge (\exists t \in \Phi ^\pi (s))(\forall s' \in (\Phi ^\pi)^{-1} (t)) \\
	& (s' \in K \wedge s' \sigma \in L(G) \Rightarrow s' \sigma \in K) \\
	& (\mbox{since } s \sigma \in K \Rightarrow s \in K \wedge s
	\sigma \in L(G))
\end{align*}

($\Leftarrow$) We have
\begin{align*}
	& s \in K \wedge s \sigma \in L(G) \\
	& \wedge (\exists t \in \Phi ^\pi (s))(\forall s' \in (\Phi ^\pi)^{-1} (t)) \\
	& (s' \in K \wedge s' \sigma \in L(G) \Rightarrow s' \sigma \in K) \\
	\Rightarrow & s \in K \wedge s \sigma \in L(G) \\
	& \wedge (s \in K \wedge s \sigma \in L(G) \Rightarrow s \sigma \in K) \\
	& (\mbox{take any $t \in \Phi ^\pi (s)$ and let $s'=s$}) \\
	\Rightarrow & s \sigma \in K
\end{align*}

Next, we show that in the definition of large language $L_a (\cS
^a/G)$, Equation (\ref{Equation9}) can be re-written as
\begin{equation} \label{Equation10}
	\begin{split}
		&  s \sigma\in L_a (\cS ^a/G) \\
		\Leftrightarrow & s \sigma \in L(G) \wedge (\sigma \in \Sigma
		_{uc} \\
		& \vee (\exists t \in \Phi ^\pi (s))(\exists \gamma \in \cS ^a(t))
		\sigma \in \gamma) \\
		\Leftrightarrow & s \sigma \in L(G) \wedge (\sigma \in \Sigma
		_{uc} \\
		& \vee (\exists t \in \Phi ^\pi (s)) \sigma \in \cS (t) \cup \Sigma _c^a )\\
		\Leftrightarrow & s \sigma \in L(G) \wedge (\sigma \in \Sigma
		_{uc} \cup \Sigma _c^a \\
		& \vee (\exists t \in \Phi ^\pi (s)) \sigma \in \cS (t)) .
	\end{split}
\end{equation}

This is because
\begin{align*}
	& (\exists \gamma \in \cS ^a(t)) \sigma \in \gamma \\
	\Leftrightarrow & (\exists \gamma \in \Delta (\cS (t))) \sigma \in \gamma \\
	& (\mbox{by Equation (\ref{Equation12})})\\
	\Leftrightarrow & (\exists \gamma ', \gamma '' \in \Sigma _c^a)
	\sigma \in (\cS (t)-\gamma ') \cup \gamma '' \\
	& (\mbox{by Equation (\ref{Equation11})})\\
	\Leftrightarrow & \sigma \in \cS (t) \cup \Sigma _c^a \\
	& (\mbox{since } \gamma '= \emptyset \wedge \gamma '' = \Sigma
	_c^a \mbox{ covers all cases}) .
\end{align*}

(IF) Assume that $K$ is CA-controllable with respect to $L(G)$,
$\Sigma _{uc}$, and $\Sigma ^a_c$ and CA-observable with respect
to $L(G)$, $\Sigma _{o}$, $\Sigma ^a_o$ and $\Phi^\pi$. We show
that $\cS_{CA}$ defined in Equation (\ref{Equation6}) is a
CA-supervisor such that $L_a(\cS^a_{CA}/G)=K$. To this end, we
prove that, for all $s \in \Sigma ^*$, $s \in L_a(\cS^a_{CA}/G)
\Leftrightarrow s \in K$ by induction on the length of $s$.

{\em Base:} Since $K$ is nonempty and closed, $\varepsilon \in K$. By
definition, $\varepsilon \in L_a(\cS^a_{CA}/G)$. Therefore, for $|s|
= 0$, that is, $s=\varepsilon$, we have
$$
s \in L_a(\cS^a_{CA}/G) \Leftrightarrow s \in K.
$$

{\em Induction Hypothesis:} Assume that for all $s \in \Sigma^*$,
$|s| \leq n$,
$$
s \in L_a(\cS^a_{CA}/G) \Leftrightarrow s \in K.
$$

{\em Induction Step:} We show that for all $s \in \Sigma^*$,
$\sigma \in \Sigma$, $|s\sigma| = n+1$,
$$
s \sigma \in L_a(\cS^a_{CA}/G) \Leftrightarrow s \sigma \in K
$$
as follows.
\begin{align*}
	& s \sigma \in L_a(\cS^a_{CA}/G) \\
	\Leftrightarrow & s \in L_a(\cS^a_{CA}/G) \wedge s
	\sigma \in L(G) \\
	& \wedge (\sigma \in \Sigma _{uc} \vee (\exists t \in \Phi ^\pi
	(s))(\exists \gamma \in \cS ^a_{CA} (t)) \sigma \in \gamma) \\
	& (\mbox{by Equation (\ref{Equation9})}) \\
	\Leftrightarrow & s \in K \wedge s \sigma \in L(G) \wedge (\sigma
	\in \Sigma _{uc} \\
	& \vee (\exists t \in \Phi ^\pi
	(s))(\exists \gamma \in \cS ^a_{CA} (t)) \sigma \in \gamma) \\
	& (\mbox{by Induction Hypothesis}) \\
	\Leftrightarrow & s \in K \wedge s \sigma \in L(G) \wedge (\sigma
	\in \Sigma _{uc} \cup \Sigma _c^a \\
	& \vee (\exists t \in \Phi ^\pi (s)) \sigma \in \cS_{CA} (t)) \\
	& (\mbox{by Equation (\ref{Equation10})}) \\
	\Leftrightarrow & (s \in K \wedge s \sigma \in L(G) \wedge \sigma
	\in \Sigma _{uc} \cup \Sigma _c^a) \\
	& \vee ( s \in K \wedge s \sigma \in L(G) \wedge (\exists t \in
	\Phi ^\pi(s)) \sigma \in \cS _{CA} (t)) \\
	\Leftrightarrow & s \sigma \in K \vee ( s \in K \wedge s \sigma
	\in L(G) \wedge (\exists t \in \Phi ^\pi(s)) \\
	& \sigma \in \cS _{CA} (t))\\
	& (\mbox{by CA-controllability of }K) \\
	\Leftrightarrow & s \sigma \in K \vee ( s \in K \wedge s \sigma
	\in L(G) \wedge (\exists t \in \Phi ^\pi(s)) \\
	& (\forall q \in  SE^\pi_H(t)) (q \in Q_H \wedge \delta
	(q,\sigma) \in Q\\
	& \Rightarrow \delta (q,\sigma) \in Q_H))\\
	& (\mbox{by the definition of } \cS _{CA} ) \\
	\Leftrightarrow & s \sigma \in K \vee ( s \in K \wedge s \sigma
	\in L(G) \wedge (\exists t \in \Phi ^\pi(s)) \\
	& (\forall s' \in (\Phi ^\pi)^{-1} (t)) (s' \in K \wedge s' \sigma
	\in L(G) \Rightarrow s' \sigma \in K))\\
	& (\mbox{by the definition of }  SE^\pi_H ) \\
	\Leftrightarrow & s \sigma \in K \vee s \sigma \in K \\
	& (\mbox{by Equation (\ref{Equation8})}) \\
	\Leftrightarrow & s \sigma \in K .
\end{align*}

(ONLY IF) Assume that there exists a CA-supervisor $\cS$ such that
$L_a(\cS^a/G)=K$. We want to prove that $K$ is CA-controllable and
CA-observable.

For CA-controllability, it can be re-written as
\begin{align*}
	(\forall s \in \Sigma ^*) (\forall \sigma \in \Sigma) & s \in K
	\wedge \sigma \in \Sigma _{uc} \cup \Sigma ^a_c \wedge s \sigma
	\in L(G) \\
	& \Rightarrow s \sigma \in K.
\end{align*}
We prove the above as follows. For all $s \in \Sigma ^*$ and
$\sigma \in \Sigma$, we have
\begin{align*}
	& s \in K \wedge \sigma \in \Sigma _{uc} \cup \Sigma ^a_c \wedge s
	\sigma \in L(G) \\
	\Rightarrow & s \in L_a(\cS^a/G) \wedge s \sigma \in L(G) \wedge
	\sigma \in \Sigma _{uc} \cup \Sigma ^a_c \\
	& (\mbox{since } L_a(\cS^a/G)=K) \\
	\Rightarrow & s \in L_a(\cS^a/G) \wedge s \sigma \in L(G) \wedge
	(\sigma \in \Sigma _{uc} \cup \Sigma _c^a \\
	& \vee (\exists t \in \Phi ^ \pi (s)) \sigma \in \cS (t)) \\
	\Rightarrow & s \sigma \in L_a(\cS^a/G) \\
	& (\mbox{by Equation (\ref{Equation10})}) \\
	\Rightarrow & s \sigma \in K \\
	& (\mbox{since } L_a(\cS^a/G)=K) .
\end{align*}

We prove CA-observability by contradiction. Suppose $K$ is
CA-controllable with respect to $L(G)$, $\Sigma _{uc}$, and
$\Sigma ^a_c$, but not CA-observable with respect to $L(G)$,
$\Sigma _{o}$, $\Sigma ^a_o$ and $\Phi^\pi$. By the definitions of
CA-observability, we have
\begin{align*}
	& (\exists s \in \Sigma ^*)(\exists \sigma \in \Sigma) s \sigma
	\in K \wedge (\forall t \in \Phi ^ \pi (s)) \\
	& (\exists s' \in (\Phi ^ \pi)^{-1} (t)) s' \in K \wedge s' \sigma
	\in L(G) \wedge s' \sigma \not\in K \\
	& (\mbox{by Equation (\ref{Equation7})}) \\
	\Leftrightarrow & (\exists s \in \Sigma ^*)(\exists \sigma \in
	\Sigma) s \sigma \in K \wedge (\forall t \in \Phi ^ \pi (s)) \\
	& (\exists s' \in \Sigma ^*) t \in \Phi ^ \pi (s') \wedge s' \in K
	\wedge s' \sigma \in L(G) \\
	&\wedge s' \sigma \not\in K \\
	& (\mbox{by the definition of } (\Phi ^ \pi)^{-1} ) .
\end{align*}
Let us consider two possible cases.

{\em Case 1}: $(\exists t' \in \Phi ^ \pi (s)) \sigma \in \cS (t')$.
In this case, by the derivation above,
\begin{align*}
	& (\exists s \in \Sigma ^*)(\exists \sigma \in \Sigma)  s \sigma
	\in K \wedge (\exists t' \in \Phi ^ \pi (s)) \sigma \in \cS (t') \\
	& \wedge (\forall t \in \Phi ^ \pi (s))(\exists s' \in \Sigma ^*) t
	\in \Phi ^ \pi (s') \\
	& \wedge s' \in K \wedge s' \sigma \in L(G) \wedge s' \sigma
	\not\in K \\
	\Rightarrow & (\exists s \in \Sigma ^*)(\exists \sigma \in
	\Sigma) (\exists t' \in \Phi ^ \pi (s)) \sigma \in \cS (t')\\
	& \wedge (\forall t \in \Phi ^ \pi (s))(\exists s' \in \Sigma ^*) t
	\in \Phi ^ \pi (s') \\
	& \wedge s' \in K \wedge s' \sigma \in L(G) \wedge s' \sigma
	\not\in K \\
	\Rightarrow & (\exists s \in \Sigma ^*)(\exists \sigma \in
	\Sigma) (\exists t' \in \Phi ^ \pi (s)) \sigma \in \cS (t')\\
	& \wedge (\exists s' \in \Sigma ^*) t' \in \Phi ^ \pi (s') \\
	& \wedge s' \in K \wedge s' \sigma \in L(G) \wedge s' \sigma
	\not\in K \\
	& (\mbox{take }t=t') \\
	\Rightarrow & (\exists s' \in \Sigma ^*)(\exists \sigma \in
	\Sigma) s' \sigma \not\in K \wedge s' \in K \\
	& \wedge s' \sigma \in L(G) \wedge (\exists t' \in \Phi ^ \pi (s'))
	\sigma \in \cS (t') \\
	\Rightarrow & (\exists s' \in \Sigma ^*)(\exists \sigma \in
	\Sigma) s' \sigma \not\in K \wedge s' \in L_a (\cS ^a/G) \\
	& \wedge s' \sigma \in L(G) \wedge (\exists t' \in \Phi ^ \pi (s'))
	\sigma \in \cS (t') \\
	& (\mbox{because } L_a(\cS ^a/G)=K) \\
	\Rightarrow & (\exists s' \in \Sigma ^*)(\exists \sigma \in
	\Sigma) s' \sigma \not\in K \wedge s' \sigma \in L_a
	(\cS^a/G) \\
	& (\mbox{by Equation (\ref{Equation10})}) .
\end{align*}
which contradicts the assumption that $L_a(\cS^a/G)=K$.

{\em Case 2}: $(\forall t' \in \Phi ^ \pi (s)) \sigma \not\in \cS
(t')$. In this case, we have
\begin{align*}
	& (\exists s \in \Sigma ^*)(\exists \sigma \in \Sigma) s \sigma
	\in K \wedge (\forall t' \in \Phi ^ \pi (s)) \sigma \not\in \cS (t') \\
	& \wedge (\forall t \in \Phi ^ \pi (s))(\exists s' \in \Sigma ^*) t
	\in \Phi ^ \pi (s') \\
	& \wedge s' \in K \wedge s' \sigma \in L(G) \wedge s' \sigma
	\not\in K \\
	\Rightarrow & (\exists s \in \Sigma ^*)(\exists \sigma \in \Sigma)
	s \sigma \in K \\
	& \wedge (\forall t' \in \Phi ^ \pi (s)) \sigma \not\in
	\cS (t') \wedge \sigma \not\in \Sigma _{uc} \cup \Sigma _c^a \\
	& (\mbox{since $K$ is CA-controllable}) \\
	\Rightarrow & (\exists s \in \Sigma ^*)(\exists \sigma \in \Sigma)
	s \sigma \in K \\
	& \wedge \neg (\sigma \in \Sigma _{uc} \cup \Sigma
	_c^a  \vee (\exists t' \in \Phi ^ \pi (s)) \sigma \in \cS (t')) \\
	\Rightarrow & (\exists s \in \Sigma ^*)(\exists \sigma \in \Sigma)
	s \sigma \in K \wedge s \sigma \not\in L_a (\cS^a/G) \\
	& (\mbox{by Equation (\ref{Equation10})}) .
\end{align*}
which contradicts the assumption that $L_a(\cS^a/G)=K$. Hence, $K$
is CA-observable with respect to $L(G)$, $\Sigma _{o}$, $\Sigma
^a_o$ and $\Phi^\pi$.

\section{PROOF OF PROPOSITION \ref{proposition 1}}	
We prove this by contradiction. Suppose
$$
\neg (((\forall j) \cS_j\in \Omega) \Rightarrow \cup \cS_{j}\in \Omega).
$$
Then
\begin{align*}
	&(\forall j) \cS_j\in \Omega \wedge \cup \cS_{j} \not\in
	\Omega \\
	\Rightarrow
	&(\forall j) L_a(\cS_j^a/G) = K \wedge L_a(\cup \cS^a_{ j}/G) \neq K\\
	&\text{(by Equation (\ref{solution}))}\\
	\Rightarrow
	&(\forall j) L_a(\cS_j^a/G) = K \wedge L_a(\cup \cS^a_{ j}/G) \supset K\\
	&\text{(by Equation (\ref{inclusion}), $K = L_a(\cS_j^a/G) \subseteq L_a(\cup \cS^a_{ j}/G)$)}\\
	\Rightarrow
	&(\forall j) L_a(\cS_j^a/G) = K \wedge (\exists s \in K)(\exists \sigma \in \Sigma) \\
	&s\sigma \notin K \wedge s\sigma \in  L_a(\cup \cS^a_{ j}/G)\\
	\Rightarrow
	&(\forall j) L_a(\cS_j^a/G) = K \wedge (\exists s \in K)(\exists \sigma \in \Sigma) \\
	&(\forall j) s\sigma \notin L_a(\cS_j^a/G) \wedge s\sigma \in  L_a(\cup \cS^a_{ j}/G)\\
	\Rightarrow
	&(\exists s \in K)(\exists \sigma \in \Sigma) (\forall j) (\forall t\in \Phi^\pi(s))\\
	& \sigma \notin \cS_j(t) \wedge s\sigma \in L_a(\cup \cS^a_{ j}/G)\\
	&\text{(by the definition of the large language)}\\
	\Rightarrow
	&(\exists s \in K)(\exists \sigma \in \Sigma) (\forall t\in \Phi^\pi(s)) (\forall j)\\
	& \sigma \notin \cS_j(t) \wedge s\sigma \in L_a(\cup \cS^a_{ j}/G)\\
	\Rightarrow
	&(\exists s \in K)(\exists \sigma \in \Sigma) (\forall t\in \Phi^\pi(s))\sigma \notin
	\cup \cS_{j}(t) \wedge \\
	&s\sigma \in L_a(\cup \cS^a_{ j}/G)\\
	&\text{(by Equation (\ref{solutioncup}))}\\
	\Rightarrow
	&(\exists s \in K)(\exists \sigma \in \Sigma) \\
	&s\sigma \notin  L_a(\cup \cS^a_{ j}/G) \wedge s\sigma \in  L_a(\cup \cS^a_{ j}/G)\\
	&\text{(by the definition of the large language)} .
\end{align*}
We hence obtain a contradiction.

\section{PROOF OF THEOREM \ref{Theorem 4}}		
We prove the theorem by contradiction. Since $\Omega \ne
\emptyset$, $\cS_{CA} \in \Omega$. Suppose $\cS_{CA}=\cS^\circ$
is not true, then
\begin{align*}
	& \cS_{CA} \not =\cS^\circ \\
	\Rightarrow & \cS_{CA}<\cS^\circ \\
	& (\mbox{because } \cS_{CA}\in \Omega \Rightarrow \cS_{CA} \leq
	\cS^\circ) \\
	\Rightarrow &(\exists t\in \Phi^\pi(L(H)))(\exists\sigma\in
	\Sigma) \sigma\in \cS^\circ(t) \wedge \sigma \notin\cS_{CA}(t) \\
	\Rightarrow &(\exists t\in \Phi^\pi(L(H)))(\exists\sigma\in
	\Sigma) \sigma\in \cS^\circ(t) \\
	& \wedge \sigma \notin (\Sigma -\eta(SE^\pi_H(t)))\cup \Sigma
	_{uc} \\
	& (\mbox{by Equation (\ref{Equation6})}) \\
	\Rightarrow &(\exists t\in \Phi^\pi(L(H)))(\exists\sigma\in
	\Sigma) \sigma\in \cS^\circ(t) \\
	& \wedge \sigma \notin (\Sigma -\eta ( SE^\pi_H(t)))\wedge
	\sigma \not\in \Sigma_{uc} \\
	\Rightarrow &(\exists t\in \Phi^\pi(L(H)))(\exists\sigma\in
	\Sigma_c) \sigma\in \cS^\circ(t) \\
	& \wedge \sigma \in \eta ( SE^\pi_H(t)) \\
	\Rightarrow &(\exists t\in \Phi^\pi(L(H)))(\exists\sigma\in
	\Sigma_c) \sigma\in \cS^\circ(t) \\
	& \wedge (\exists q \in SE^\pi_H(t)) \delta(q,\sigma)\in Q-Q_H \\
	& (\mbox{by Equation (\ref{gamma})}) \\
	\Rightarrow &(\exists t\in \Phi^\pi(L(H)))(\exists\sigma\in
	\Sigma_c) \sigma\in \cS^\circ(t) \\
	& \wedge (\exists s \in L(H)) t \in \Phi ^\pi (s) \wedge \delta _H
	(q_0,s \sigma ) \in Q-Q_H \\
	& (\mbox{by Equation (\ref{SEH})}) \\
	\Rightarrow &(\exists t\in \Phi^\pi(L(H)))(\exists \sigma \in
	\Sigma_c) \sigma\in \cS^\circ(t) \\
	& \wedge (\exists s \in L(H)) t \in \Phi ^\pi (s) \wedge s \sigma
	\in L(G) \wedge s \sigma \not\in L(H) \\
	\Rightarrow &(\exists t\in \Phi^\pi(L(H)))(\exists \sigma \in
	\Sigma_c) (\exists s \in L_a((\cS^\circ)^a/G)) \\
	& \sigma\in \cS^\circ(t) \wedge  t \in \Phi ^\pi
	(s) \wedge s \sigma \in L(G) \wedge s \sigma \not\in L(H) \\
	&\text{(because $L_a((\cS^\circ)^a/G)=L(H)=K$)} \\
	\Rightarrow &(\exists t\in \Phi^\pi(L(H)))(\exists \sigma \in
	\Sigma_c) (\exists s \in L_a((\cS^\circ)^a/G)) \\
	& t \in \Phi ^\pi(s) \wedge (\exists \gamma_a = \cS^\circ
	(t)\in (\cS^\circ) ^a(t)) \sigma \in \gamma_a \\
	& \wedge s \sigma \in L(G) \wedge s \sigma \not\in L(H) \\
	\Rightarrow &(\exists t\in \Phi^\pi(L(H)))(\exists \sigma \in
	\Sigma_c) (\exists s \in L_a((\cS^\circ)^a/G)) \\
	& s \sigma \in L_a((\cS^\circ)^a/G)) \wedge s \sigma \not\in
	L(H) \\
	&\text{(by the definition of the large language)}\\
	\Rightarrow & L_a((\cS^\circ)^a/G)) \not= L(H)=K ,
\end{align*}
which contradicts the fact $L_a((\cS^\circ)^a/G) = K$.

\section{PROOF OF PROPOSITION \ref{proposition 2}}

From the definition of parallel composition, we know that, for any string $s\in
L(G)$,
\begin{equation}\label{G||SA}
	\tilde{\delta}(y_0,s)= \tilde{\delta}((q_0, z_0),s)
	= (\delta(q_0,s),\delta_{SA}(z_0,P(s))) .
\end{equation}

Let $s=\sigma_1\sigma_2\cdots\sigma_{|s|} \in L(\tilde{G})$ and
$y_k = (q_k, z_k) = \delta(y_0,\sigma _1\cdots\sigma _{k}), k= 1,
2, ..., |s|$. Then,
\begin{equation} \label{Equation20}
	\begin{split}
		\Phi^\omega(P(s))=&\Phi ^\omega(P(\sigma_1)P(\sigma_2)\cdots P(\sigma_{|s|}))\\
		=& \omega(z_0,P(\sigma_1)) \omega(\delta_{SA}(z_0,P(\sigma_1)),
		P(\sigma_2)) \cdots \\
		&\omega(\delta_{SA}(z_0,P(\sigma_1\cdots\sigma_{{|s|}-1})),P(\sigma_{{|s|}}))
	\end{split}
\end{equation}

On the other hand, for each event $\sigma_i$, let us calculate $
P(L_i)$ as follows.

Case 1. $\sigma_i\in \Sigma_{uo}$.
\begin{align*}
	P(L_i)&=P(\{\sigma_i\})=\{\varepsilon\}\\
	&=\omega(\delta_{SA}(z_0,P(\sigma_1\cdots\sigma_{i-1})),P(\sigma_{i})) \\
	& (\mbox{since } \omega(z, \varepsilon) =\{\varepsilon\})
\end{align*}

Case 2. $\sigma_i\in \Sigma_{o}-\Sigma_{o}^a$.
\begin{align*}
	P(L_i)&=P(\{\sigma_i\})=\{\sigma_i\}\\
	&=\omega(\delta_{SA}(z_0,P(\sigma_1\cdots\sigma_{i-1})),P(\sigma_{i})) \\
	& (\mbox{since } \omega(z, \sigma)=\{\sigma\})
\end{align*}

Case 3. $\sigma_i\in \Sigma_{o}^a$.
\begin{equation*}
	\begin{aligned}
		P(L_i)=&P(\tilde{\pi}((q_{i-1}, z_{i-1}),\sigma _i, (q_{i}, z_{i})))\\
		=&\tilde{\pi}((q_{i-1}, z_{i-1}),\sigma _i, (q_{i}, z_{i}))\\
		=&\omega(z_{i-1}, \sigma_i)\\
		&\mbox{(by the definition of $\tilde{\pi}$)}\\
		=&\omega(\delta_{SA}(z_0,P(\sigma_1\cdots\sigma_{i-1})), \sigma_i)\\
		&\mbox{(by Equation (\ref{G||SA}))}\\
		=&\omega(\delta_{SA}(z_0,P(\sigma_1\cdots\sigma_{i-1})), P(\sigma_i))
	\end{aligned}
\end{equation*}

Hence, for any event $\sigma_i$ in $s$, we always have
$$
P(L_i)=\omega(\delta_{SA}(z_0,P(\sigma_1\cdots\sigma_{i-1})), P(\sigma_i)).
$$

Therefore, by Equations (\ref{tPhi}) and (\ref{Equation20}), we
have
$$
\tilde{\Phi}^{\tilde{\pi}}(s)=\Phi^\omega(P(s)).
$$

\section{PROOF OF THEOREM \ref{theorem 5}}	

For any state $q\in Q$, we have
\begin{align*}
	& q \in R(SE^{\tilde{\pi}}_{\tilde{G}}(t)) \\
	\Leftrightarrow & (\exists z \in Z) (q,z) \in
	SE^{\tilde{\pi}}_{\tilde{G}}(t) \\
	&\mbox{(by Equation (\ref{RSEt}))}\\
	\Leftrightarrow & (\exists s \in L(\tilde{G})) t \in \tilde{\Phi}
	^{\tilde{\pi}} (s) \wedge (\exists z \in Z) (q,z) = \tilde{\delta}
	(y_0, s) \\
	&\mbox{(by Equation (\ref{SEt}))}\\
	\Leftrightarrow & (\exists s \in L(\tilde{G})) t \in \tilde{\Phi}
	^{\tilde{\pi}} (s) \wedge (\exists z \in Z) \\
	& (q,z) = (\delta(q_0,s),\delta_{SA}(z_0,P(s))) \\
	&\mbox{(by Equation (\ref{G||SA}))}\\
	\Leftrightarrow & (\exists s \in L(\tilde{G})) t \in \tilde{\Phi}
	^{\tilde{\pi}} (s) \wedge q = \delta(q_0,s) \\
	&\mbox{(because $P(L(G)) \subseteq L(SA)$ implies} \\
	&\mbox{$z=\delta_{SA}(z_0,P(s))$ always exists)} \\
	\Leftrightarrow & (\exists s\in L(G)) t\in \Phi ^\omega (P(s))
	\wedge \delta(q_0 ,s)=q \\
	&\mbox{(by Proposition \ref{proposition 2})}\\
	\Leftrightarrow & q\in SE^\omega_G(t) \\
	&\mbox{(by Equation (\ref{SEGt}))} .
\end{align*}

\section{PROOF OF THEOREM \ref{theoreme6}}	
The proof is similar to the proof of Theorem \ref{theoreme3}, with
$\Phi^\pi$ and $\cS_{CA}$ replaced by $\Phi^\omega$ and
$\tilde{\cS}_{CA}$, respectively.

\section{Explanations OF ALL SYMBOLS}
Explanations of all symbols is shown in Table \ref{t of meaning of symbols}.
\begin{table*}\label{t of meaning of symbols}
	\begin{center}
		\setlength{\abovecaptionskip}{0cm}  
		\setlength{\belowcaptionskip}{-0.4cm} 
		\caption{Meaning of all symbols}
		\begin{tabular}{|c|c|c|c|}
			\hline
			Symbol& Meaning&Symbol& Meaning\\
			\hline
			     $G$& a deterministic automaton &$Q$&a set of states\\
			       $q$&a state in $Q$&$\Sigma$&a set of events\\
			    $ \sigma $&an event in $ \Sigma $&$\delta$& a (partial) transition function\\
			      $q_0$& the initial state in $G$ & $\Sigma^*$ & a set of all string over $\Sigma$\\
			       $K$& a specification language &$s$ or $t$&a string in $\Sigma^*$\\
			       $s'$& the prefix of $s$&$\overline{K}$& the prefix of $K$\\
			      $L(G)$ &the language of $G$ & $|s|$ &the length of $s$ \\
			 $|x|$&the cardinality of set $x$&$\Sigma_c$&the set of controllable events \\
			  $\Sigma_{uc}$& the set of uncontrollbale events & $\Sigma_{o}$& the set of observable events\\
			  $\Sigma_{uo}$ & the set of unobservable events&$\delta_o$ &the set of observable transitions\\
			  $\delta_{uo}$& the set of unobservable transitions& $\varepsilon$ & a empty string \\
			       $P$  &the natural mapping &$\Sigma_o^a$&  the set of attacked observable events \\
			    $\delta^a$&the set of attacked transitions & $tr$&  a transition in $\delta^a$ \\
			     $A_{tr}$ &  the subset of $\Sigma_o^*$  & $\mathbb{A}$ &  the set of $A_{tr}$ \\
			      $\pi$&  the mapping from $\delta^a$ to $ \mathbb{A} $ &  $ 2^{\Sigma} $  & the power set of $ \Sigma $  \\
			  $ \Theta^\pi $  &   the mapping from $ L(G) $ to $ 2^{\Sigma^*} $  &  $ \Phi^\pi $ & the mapping  consisted of $ P $ and $ \Theta^\pi $ \\
			 $ \mathcal{S} $ &  a supervisor& $ \Gamma $&the set of all possible controls\\
			  $ \Sigma_c^a $ & the set of attacked controllable events &$ \gamma $ &  a control  \\
			$ \Delta(\gamma) $ & the set of possible control for $ \gamma $ after attacks & $ \mathcal{S}^a $ & the attacked supervisor\\
$ \alpha, \beta, \lambda, \mu $ & events is examples &$ \mathcal{S}^a/G $ &the supervised system under actuator attacks\\
$L_a(\mathcal{S}^a/G)$&the large language generated by $\mathcal{S}^a/G$&$ \gamma_a $ & a possible control in $ \Delta(\gamma) $ \\			
$ H $  & a sub-automaton of $G$ &$ SE^\pi_G(t) $ &the set of state estimations of $G$ after observing $t\in\Phi^\pi(L(G))$ \\
$ F_{tr} $ & an automaton marking $ A_{tr} $ &$ G^\diamond $ &the automaton after replacing all transitions subject to attacks \\
$ \hat{Q} $ &the set of states added during the replacement&$ G^\diamond_\varepsilon $ &the automaton after replacing unobservable transitions by $\varepsilon$-transitions in $ G^\diamond $ \\
$ G^\diamond_{obs} $ &CA-observer on $G^\diamond_\varepsilon$  &$ AC(.) $ &the accessible part \\
$ UR(.) $ &the unobservable reach &$ \xi $ &the trasnition function of $ G^\diamond_{obs} $ \\
$ X_m $ &the set of marked states of $ G^\diamond_{obs} $&$ S_{CA} $ &a state-estimate-based supervisor\\
 $ (\Phi^\pi)^{-1}$&the inverse mapping of $\Phi^\pi  $ &$ \cup \cS_{j} $ &the union of a set of supervisors $S_j$ \\
$ \Omega $ &the set of all supervisors that solve SCPDES-CA &$ \cS^\circ $ &the maximally-permissive supervisor that solves SCPDES-CA\\
$ \varpi $ &the mapping of observation-based sensor attack  &$ SA $ & an automaton describing sensor attack strategy \\
$ \omega $ &the mapping of state-based sensor attack  &$ \Phi^\omega $ &the mapping of the possible observations under $ \omega $ \\
$ \tilde{G} $&$ G||SA$ and all symblos with $ \sim$ is w.r.t $\tilde{G}$ &$ SE^\omega_G(t) $ &the set of state estimations of $G$ after observing $t\in\Phi^\omega(L(G))$\\
	\hline
		\end{tabular}
	\end{center}
\end{table*}

\bibliographystyle{IEEEtran}

\bibliography{CyberRefe}

\end{document}